\documentclass[journal]{IEEEtran}
\usepackage{ifpdf}
\ifpdf
    \usepackage[pdftex]{graphicx}
    \usepackage[update]{epstopdf}
\else
	\usepackage{graphicx}
\fi

\usepackage{epsfig,amsmath,amssymb,amsfonts,amstext,amsthm,enumerate}
\usepackage{latexsym,graphics,epsf,epsfig,color,float}
\usepackage{cite,url}
\usepackage{setspace}
\usepackage{multirow}
\usepackage{pdflscape, dashrule}
\usepackage{bbm}
\newtheorem{thm}{Theorem}

\newtheorem{lemma}{Lemma}

\allowdisplaybreaks

\begin{document}

\title{On the Tradeoff Region of Secure Exact-Repair Regenerating Codes}
\author{Shuo Shao, Tie Liu, Chao Tian, and Cong Shen
\thanks{This work was supported in part by the National Science Foundation under Grants CCF-13-20237, CCF-15-24839, and CCF-15-26095, and in part by the National Science Foundation of China under Grant 61631017.}
\thanks{S.~Shao and T.~Liu are with the Department of Electrical and Computer Engineering, Texas A\&M University, College Station, TX 77843, USA (e-mail: \{shaoshuo,tieliu\}@tamu.edu).}
\thanks{C.~Tian is with the Department of Electrical Engineering and Computer Science, University of Tennessee, Knoxville, TN 37996, USA (e-mail: ctian1@utk.edu).}
\thanks{C.~Shen is with the Department of Electronic Engineering and Information Science, University of Science and Technology of China, Hefei, Anhui 230027, China (e-mail: congshen@ustc.edu.cn).}}

\maketitle

\begin{abstract}
We consider the $(n,k,d,\ell)$ secure exact-repair regenerating code problem, which generalizes the $(n,k,d)$ exact-repair regenerating code problem with the additional constraint that the stored file needs to be kept information-theoretically secure against an eavesdropper, who can access the data transmitted to regenerate a total of $\ell$ different failed nodes. For all known results on this problem, the achievable tradeoff regions between the normalized storage capacity and repair bandwidth have a single corner point, achieved by a scheme proposed by Shah, Rashmi and Kumar (the SRK point). Since the achievable tradeoff regions of the exact-repair regenerating code problem without any secrecy constraints are known to have multiple corner points in general, these existing results suggest a phase-change-like behavior, i.e., enforcing a secrecy constraint ($\ell\geq 1$) immediately reduces the tradeoff region to one with a single corner point. In this work, we first show that when the secrecy parameter $\ell$ is sufficiently large, the SRK point is indeed the only corner point of the tradeoff region. However, when $\ell$ is small, we show that the tradeoff region can in fact have multiple corner points. In particular, we establish a precise characterization of the tradeoff region for the $(7,6,6,1)$ problem, which has exactly two corner points. Thus, a smooth transition, instead of a phase-change-type of transition, should be expected as the secrecy constraint is gradually strengthened.
\end{abstract}


% \fi
%\IEEEpeerreviewmaketitle


\section{Introduction}
Fault tolerance and node repair are two fundamental ingredients of reliable distributed storage systems. While the study of fault tolerance via diversity coding has been in the literature for decades \cite{Sin-IT64,Roc-Thesis92,Yeu-IT95,RYH-IT97,YZ-IT99,MTD-IT10,JML-IT14}, systematic studies of node repair mechanisms were started only recently by Dimakis et al. in their pioneering work \cite{Dimakis-IT10}. A particular model, which has received a significant amount of attention in the literature, is the so-called {\em exact-repair regenerating code} problem.

\begin{figure}
\centering
\includegraphics[width=0.9\linewidth,draft=false]{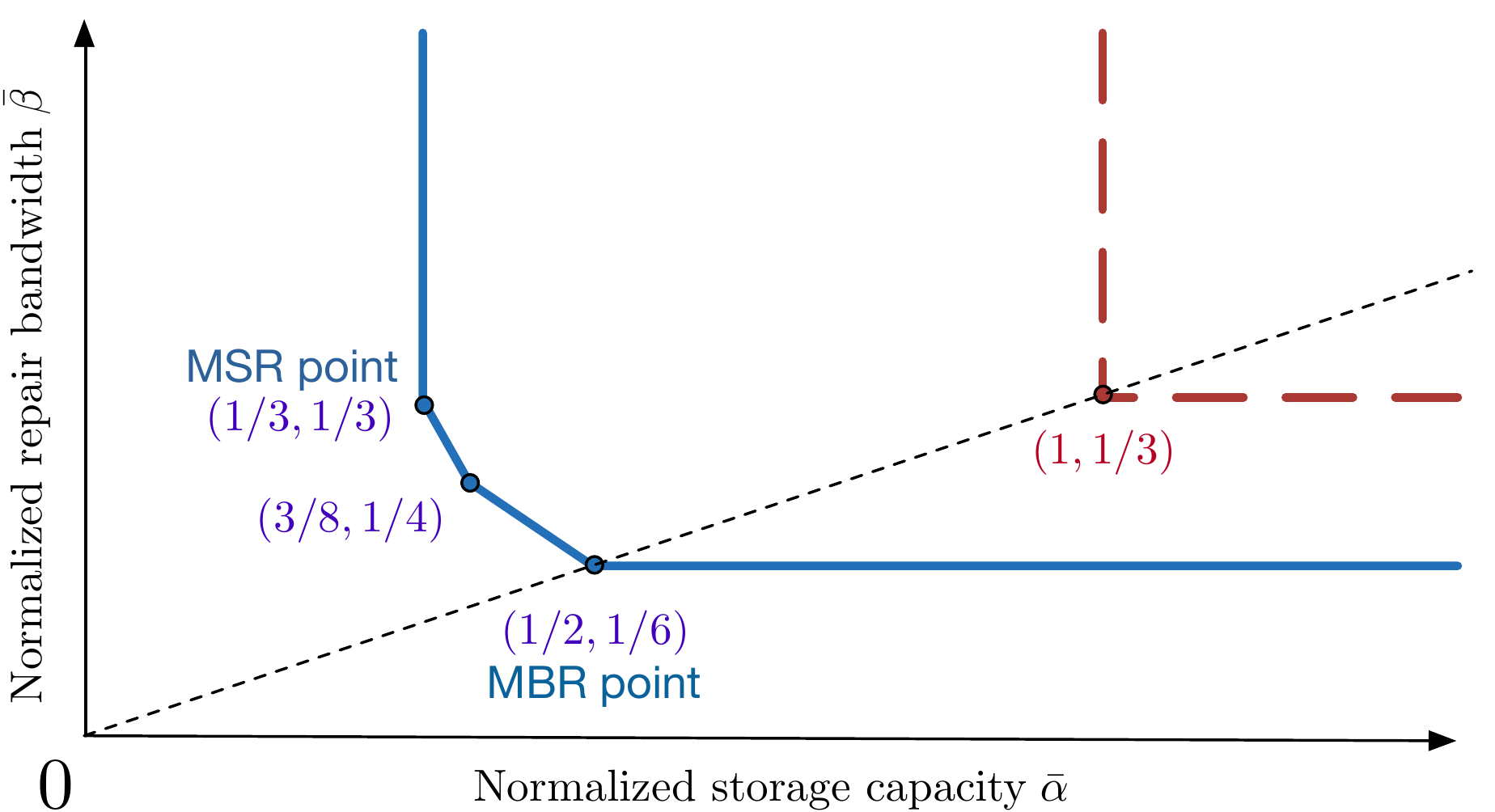}
\caption{The regions above the solid and the dashed lines are the achievable normalized storage-capacity repair-bandwidth tradeoff regions for the $(4,3,3)$ exact-repair regenerating code and the $(4,3,3,1)$ secure exact-repair regenerating code problems, respectively.}
\label{fig}
\end{figure}

More specifically, in an $(n,k,d)$ exact-repair regenerating code problem, a file $\mathsf{M}$ of size $B$ is to be encoded and then stored in a total of $n$ distributed storage nodes, each of capacity $\alpha$. The encoding needs to ensure that: 1) the file $\mathsf{M}$ can be perfectly recovered by having {\em full} access to any $k$ out of the total $n$ storage nodes; 2) when a node failure occurs, the failed node can be regenerated by {\em extracting} data of size $\beta$ from each of an arbitrary set of $d$ remaining nodes. An important technical contribution of \cite{Dimakis-IT10} was to show that there is an inherent {\em tradeoff} between the node capacity $\alpha$ and the repair bandwidth $\beta$ in satisfying both the file-recovery and node-regeneration requirements. In particular, it has been shown \cite{Kumar-IT11} that the achievable normalized storage-capacity repair-bandwidth tradeoff region for any $(n,k,d)$ exact-repair regenerating code problem with $k>1$ features {\em multiple} corner points including the the all-important {\em minimum storage rate (MSR)} and {\em minimum bandwidth rate (MBR)} points. Fig.~\ref{fig} illustrates the achievable normalized storage-capacity repair-bandwidth tradeoff region for the $(4,3,3)$ exact-repair regenerating code problem, which features three corner points including the MSR point $(1/3,1/3)$ and the MBR point $(1/2,1/6)$. Despite intensive research efforts that have yielded many highly non-trivial partial results \cite{Dimakis-IT10,Kumar-IT11,Tian-JSAC13,Duursma-P14}, the optimal tradeoffs between the node capacity $\alpha$ and repair bandwidth $\beta$ have {\em not} been fully understood for the general $(n,k,d)$ exact-repair regenerating code problem.

In this paper, we consider an extension of the aforementioned exact-repair regenerating code problem, which further requires certain security guarantee during the node-regeneration processes. More specifically, the $(n,k,d,\ell)$ {\em secure exact-repair regenerating code} problem that we consider is the standard $(n,k,d)$ exact-repair regenerating code problem \cite{Dimakis-IT10,Kumar-IT11,Tian-JSAC13,Duursma-P14}, with the additional constraint that the file $\mathsf{M}$ needs to be kept {\em information-theoretically} secure against an eavesdropper that can access the data extracted to regenerate a total of $\ell$ different failed nodes (possibly under different repair groups). Apparently, this is only possible when $\ell<k$. Furthermore, when $\ell=0$, the secrecy constraint degenerates, and the $(n,k,d,\ell)$ secure exact-repair regenerating code problem reduces to the $(n,k,d)$ exact-repair regenerating code problem without any security constraints.

Under the additional secrecy constraint ($\ell \geq 1$), the optimal tradeoffs between the node capacity $\alpha$ and repair bandwidth $\beta$ have been studied in \cite{PRR-ISIT10,SRK-Globecom11,TACB-IT16,YSY-ISIT16}. In particular, Shah, Rashmi and Kumar \cite{SRK-Globecom11} showed that a particular tradeoff point (referred to as the {\em SRK} point) can be obtained by extending an MBR code based on the product-matrix construction proposed in \cite{Kumar-IT11}. Later, it was shown that the SRK point is the {\em only} corner point of the tradeoff region for the cases where we have either $d=2,3$ \cite{TACB-IT16}, or $d=4$ \cite{YSY-ISIT16}, or $k=2$ \cite{TACB-IT16}, or $\ell=k-1=d-1$ \cite{TACB-IT16}. This is in sharp contrast to the original exact-repair regenerating code problem \cite{Dimakis-IT10,Kumar-IT11,Tian-JSAC13,Duursma-P14} without any secrecy constraints, for which, as mentioned previously, the tradeoff region features {\em multiple} corner points when $k>1$. Fig.~\ref{fig} also illustrates the tradeoff region for the $(4,3,3,1)$ secure exact-repair regenerating code problem, which features a single corner point at $(1,1/3)$. Thus, the existing results from \cite{TACB-IT16,YSY-ISIT16} seem to suggest a {\em phase-change-like} behavior that enforcing a secrecy constraint  immediately reduces the tradeoff region from one with multiple corner points ($\ell=0$) to one with a single corner point ($\ell\geq 1$). 

\begin{figure}
\centering
\includegraphics[width=0.9\linewidth,draft=false]{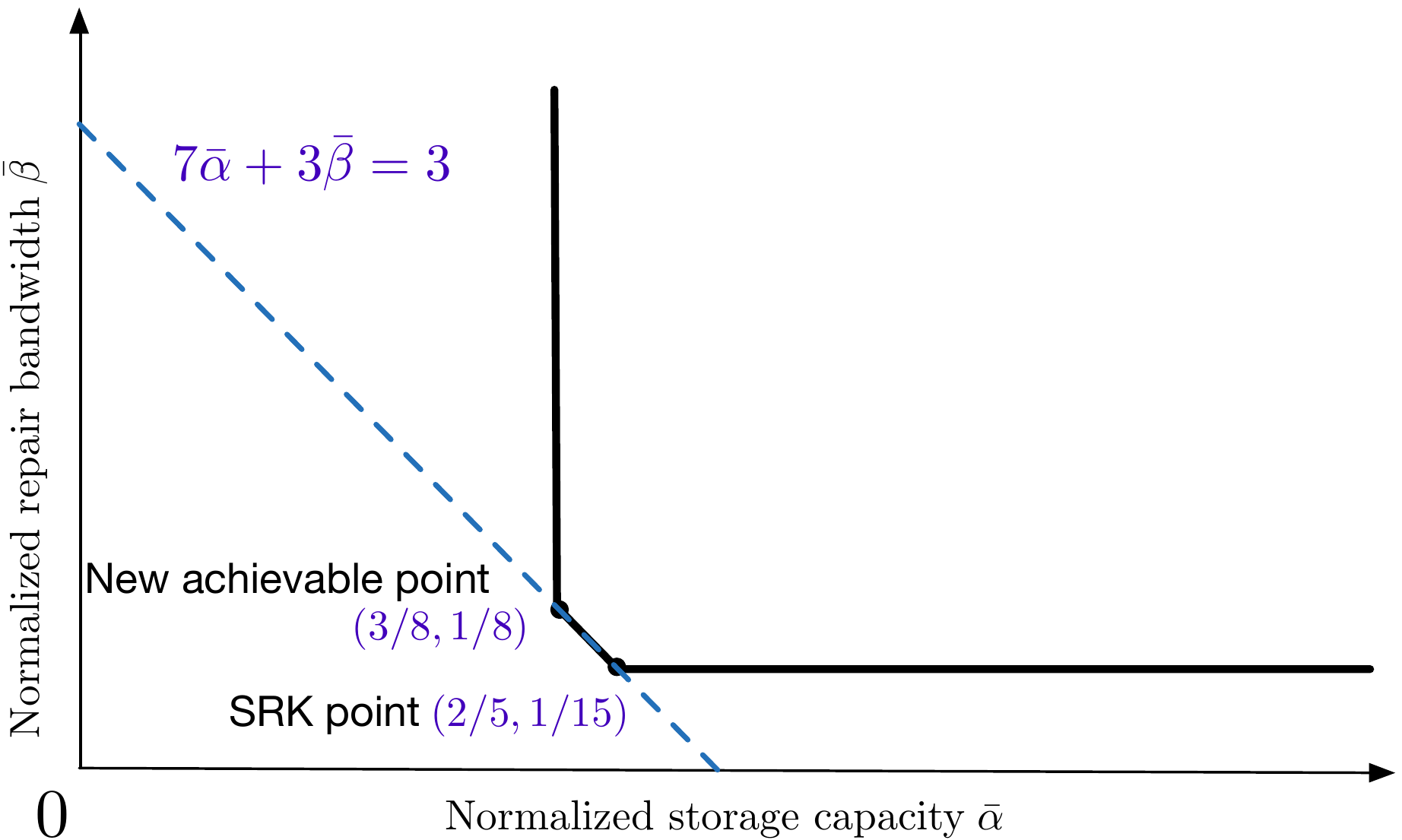}
\caption{The regions above the solid line is the achievable normalized storage-capacity repair-bandwidth tradeoff region for the $(7,6,6,1)$ secure exact-repair regenerating code problem. In addition to the SRK point $(2/15,1/5)$, the tradeoff region has another corner point at $(3/8,1/8)$.}
\label{fig1}
\end{figure}

The main results of this paper are two-folded.
\begin{itemize} 
\item We first show, via new converse results, that for any given $(k,d)$ pair, there is a lower bound on $\ell$, denoted by $\ell^*(k,d)$, such that when $\ell \geq \ell^*(k,d)$, the SRK point is indeed the {\em only} conner point of the tradeoff region for the $(n,k,d,\ell)$ secure exact-repair regenerating code problem. As we shall see, the lower bound $\ell^*(k,d)\leq k-1$ for any $(k,d)$ pair, and thus the tradeoff region for any $(n,k,d,\ell)$ problem with $\ell= k-1$ or $k=2$ must have a single corner point. In addition, the lower bound $\ell^*(k,d)=1$ for any $d\in[2:4]$. Therefore, our result includes all previous results from \cite{TACB-IT16} and \cite{YSY-ISIT16} as special cases. 
\item Next, we show that when $1 \leq \ell < \ell^*(k,d)$, it is entirely possible that the tradeoff region features {\em multiple} corner points. In particular, we establish a precise characterization of the tradeoff region for the $(7,6,6,1)$ problem, which has exactly {\em two} corner points (see Fig.~\ref{fig1} for an illustration). This result requires new achievability results as well as new converse results, the former of which are obtained by extending the layered coding scheme proposed in \cite{TSAVK-IT15}. From the viewpoint of the rate region, our result suggests that a {\em smooth} transition, instead of a phase-change-type of transition, should be expected as the secrecy constraint is gradually strengthened by increasing the parameter $\ell$.
\end{itemize}

%the transition from the original exact-repair regenerating code problem to the secrecy extension is {\em not} an ``abrupt" one as suggested by the previous results from \cite{TACB-IT16} and \cite{YSY-ISIT16}.

\section{Problem Formulation and Known Results}
Let $(n,k,d,N,K,T,S)$ be a tuple of positive integers such that $n \geq d+1 \geq k+1 \geq 2$. Formally, an $(n,k,d,N,K,T,S)$ code consists of:
\begin{itemize}
\item for each $i \in [1:n]$, a {\em message-encoding} function $f_i: [1:N]\times[1:K] \rightarrow [1:T]$;
\item for each $\mathcal{A} \subseteq [1:n]$ such that $|\mathcal{A}|=k$, a {\em message-decoding} function $g_\mathcal{A}: [1:T]^k \rightarrow [1:N]$;
\item for each $\mathcal{B}\subseteq [1:n]$ such that $|\mathcal{B}|=d$, $i \in \mathcal{B}$, and $j \in [1:n]\setminus\mathcal{B}$, a {\em repair-encoding} function $f^{\mathcal{B}}_{i \rightarrow j}: [1:T] \rightarrow [1:S]$;
\item for each $\mathcal{B}\subseteq [1:n]$ such that $|\mathcal{B}|=d$ and $j \in [1:n]\setminus\mathcal{B}$, a {\em repair-decoding} function $g^{\mathcal{B}}_j:[1:S]^d \rightarrow [1:T]$. 
\end{itemize} 

Let $\mathsf{M}$ be a message that is uniformly distributed over $[1:N]$, and $\mathsf{K}$ be a secret key that is uniformly distributed over $[1:K]$. The message $\mathsf{M}$ and the secret key $\mathsf{K}$ are assumed to be independent of each other. For each $i\in[1:n]$, let $\mathsf{W}_i=f_i(\mathsf{M},\mathsf{K})$ be the data stored at the $i$th storage node, and for each $\mathcal{B}\subseteq [1:n]$ such that $|\mathcal{B}|=d$, $i \in \mathcal{B}$, and $j \in [1:n]\setminus\mathcal{B}$, let $\mathsf{S}^{\mathcal{B}}_{i\rightarrow j}=f^{\mathcal{B}}_{i\rightarrow j}(\mathsf{W}_i)$ be the data extracted from the $i$th storage node in order to regenerate the data stored at the $j$th storage node under the context of repair group $\mathcal{B}$. Obviously, 
\begin{align*}
B&=\log{N}, \quad \alpha=\log{T}, \quad \beta=\log{S}
\end{align*}
represent the message rate, storage capacity, and repair bandwidth, respectively.

A normalized storage-capacity repair-bandwidth pair $(\bar{\alpha},\bar{\beta})$ is said to be {\em achievable} for the $(n,k,d,\ell)$ secure exact-repair regenerating code problem if an $(n,k,d,N,K,T,S)$ code can be found such that:
\begin{itemize}
\item (rate normalization) 
\begin{align}
{\alpha}/{B}=\bar{\alpha}  \quad \mbox{and} \quad {\beta}/{B}=\bar{\beta};\label{eq:Cons1}
\end{align}
\item (message recovery) 
\begin{align}
\mathsf{M} =g_\mathcal{A}(\mathsf{W}_i:i\in\mathcal{A})\label{eq:Cons2}
\end{align}
for any $\mathcal{A} \subseteq [1:n]$ such that $|\mathcal{A}|=k$;
\item (node regeneration) 
\begin{align}
\mathsf{W}_j =g^{\mathcal{B}}_j(\mathsf{S}^{\mathcal{B}}_{i\rightarrow j}:i\in\mathcal{B})\label{eq:Cons3}
\end{align}
for any $\mathcal{B}\subseteq [1:n]$ such that $|\mathcal{B}|=d$ and $j \in [1:n]\setminus\mathcal{B}$;
\item (repair secrecy) 
\begin{align}
I(\mathsf{M};(\mathsf{S}_{\rightarrow j}:j\in \mathcal{E}))=0\label{eq:Cons4}
\end{align}
for any $\mathcal{E}\subseteq [1:n]$ such that $|\mathcal{E}|=\ell$, where $\mathsf{S}_{\rightarrow j} :=(\mathsf{S}^{\mathcal{B}}_{i\rightarrow j}:\mathcal{B}\subseteq [1:n], \; |\mathcal{B}|=d, \; \mathcal{B}\not\ni j, \; i \in \mathcal{B})$ is the collection of data that can be extracted from the other nodes to regenerate node $j$.
\end{itemize}
The closure of all achievable $(\bar{\alpha},\bar{\beta})$ pairs is the {\em achievable normalized storage-capacity repair-bandwidth tradeoff region} $\mathcal{R}_{n,k,d,\ell}$ for the $(n,k,d,\ell)$ secure exact-repair regenerating code problem.

In \cite{SRK-Globecom11}, Shah, Rashmi and Kumar proved the following important achievability result for the general $(n,k,d,\ell)$ secure regenerating code problem:
\begin{align}
\left(dT_{k,d,\ell,}^{-1},T_{k,d,\ell}^{-1}\right)\in \mathcal{R}_{n,k,d,\ell}\label{eq:MBR2}
\end{align}
where
\begin{align}
T_{k,d,\ell}:=\sum_{i=\ell+1}^{k}(d+1-i).
\end{align}
Note that when $\ell=0$ (no repair-secrecy constraint), $\left(dT_{k,d,0}^{-1},T_{k,d,0}^{-1}\right)$ recovers the MBR point of the $(n,k,d)$ exact-repair regenerating code problem \cite{Kumar-IT11}. It has been shown that the SRK point \eqref{eq:MBR2} is the {\em only} corner point of the tradeoff region $\mathcal{R}_{n,k,d,\ell}$ for the cases where we have either $d=2,3$ \cite{TACB-IT16}, or $d=4$ \cite{YSY-ISIT16}, or $k=2$ \cite{TACB-IT16}, or $\ell=k-1=d-1$ \cite{TACB-IT16}. 

\section{New Results}
Consider the $(n,k,d,\ell)$ secure exact-repair regenerating code problem (with $\ell \geq 1$), and let
\begin{align}
\ell^*(k,d):=\min\left\{\ell\geq1:T_{k,d,\ell} \leq d+\sqrt{d\ell}\right\}.
\end{align}
Note that $T_{k,d,\ell}$ is monotone non-increasing with respect to $\ell$ for any given $(k,d)$ pair, so we have
\begin{align}
T_{k,d,\ell} \leq d+\sqrt{d\ell}, \quad \forall \ell \geq \ell^*(k,d). \label{eq:Cond}
\end{align}
We have the following two {\em outer} bounds for the tradeoff region $\mathcal{R}_{n,k,d,\ell}$.

\begin{thm}\label{thm}
For the general $(n,k,d,\ell)$ secure exact-repair regenerating code problem, any achievable normalized storage-capacity repair-bandwidth pair $(\bar{\alpha},\bar{\beta}) \in \mathcal{R}_{n,k,d,\ell}$ must satisfy:
\begin{align}
\bar{\beta} & \geq T_{k,d,\ell}^{-1}.\label{eq:C1}
\end{align}
In addition, when $\ell \geq \ell^*(k,d)$, any achievable normalized storage-capacity repair-bandwidth pair $(\bar{\alpha},\bar{\beta}) \in \mathcal{R}_{n,k,d,\ell}$ must also satisfy:
\begin{align}
\bar{\alpha} & \geq dT_{k,d,\ell}^{-1}.\label{eq:C2}
\end{align}
(Conversely, any $(\bar{\alpha},\bar{\beta})$ satisfying (\ref{eq:C1}) and (\ref{eq:C2}) is achievable.)
\end{thm}

While the proof of \eqref{eq:C1} is straightforward, the proof of \eqref{eq:C2} is long and technical. We shall defer the proof to Section~\ref{sec:Converse}. Combining \eqref{eq:C1} and \eqref{eq:C2} proves that the SRK point \eqref{eq:MBR2} is the {\em only} corner point of the tradeoff region $\mathcal{R}_{n,k,d,\ell}$ when $\ell \geq \ell^*(k,d)$. It is straightforward to verify that the lower bound $\ell^*(k,d)\leq k-1$ for any $(k,d)$ pair and $\ell^*(k,d)=1$ for $d\in[2:4]$. Therefore, Theorem~\ref{thm} includes all previous results from \cite{TACB-IT16} and \cite{YSY-ISIT16} as special cases.

Next, we shift our attention to the cases where $1 \leq \ell < \ell^*(k,d)$. To see how the tradeoff region $\mathcal{R}_{n,k,d,\ell}$ may look like in this case, let us begin with the following achievability results for the $(n,k,d,\ell)$ secure exact-repair regenerating code problem with $k=d=n-1$.

\begin{thm}\label{thm2}
For any $t\in [2:n-\ell]$, we have
\begin{align}
(\bar{\alpha}_t,\bar{\beta}_t) \in \mathcal{R}_{n,n-1,n-1,\ell}
\end{align}
where
\begin{align}
(t-1)\bar{\alpha}_t=(n-1)\bar{\beta}_t
:=\left.
\left(
\begin{array}{c}
  n-1   \\
  t-1   
\end{array}
\right)
\right/
{
\left(
\begin{array}{c}
  n-\ell   \\
  t   
\end{array}
\right)
}.\label{eq:NTP}
\end{align}
\end{thm}

The proof is based on a new coding scheme, which we shall describe in the next section. Note that when $\ell=1$, $(\bar{\alpha}_t,\bar{\beta}_t)$ can be simplified as:
\begin{align}
(\bar{\alpha}_t,\bar{\beta}_t)=\left(\frac{t}{(t-1)(n-t)},\frac{t}{(n-1)(n-t)}\right).
\end{align}
In this case, when $t=2$, $(\bar{\alpha}_t,\bar{\beta}_t)$ coincides with the SRK point \eqref{eq:MBR2} with $k=d=n-1$ and $\ell=1$. Furthermore, note that $\bar{\beta}_t$ is monotone increasing with $t$, and $\bar{\alpha}_t$ is monotone decreasing with $t$ for any $t\in[2:n-1]$ such that $t^2+t<n$. Thus, {\em no} pairs of points from the set $\{(\bar{\alpha}_2,\bar{\beta}_2),\ldots,(\bar{\alpha}_{t+1},\bar{\beta}_{t+1})\}$ dominate each other for any $t\in[2:n-1]$ such that $t^2+t<n$. For example, when $n=7$, a second achievability point $(\bar{\alpha}_3,\bar{\beta}_3)=\left(\frac{3}{8},\frac{1}{8}\right)$ emerges in addition to the SKR point $(\bar{\alpha}_2,\bar{\beta}_2)=\left(\frac{2}{5},\frac{1}{15}\right)$. When $n=13$, a third achievability point $(\bar{\alpha}_4,\bar{\beta}_4)=\left(\frac{4}{27},\frac{1}{27}\right)$ emerges in addition to the points $(\bar{\alpha}_3,\bar{\beta}_3)=\left(\frac{3}{20},\frac{1}{40}\right)$ and $(\bar{\alpha}_2,\bar{\beta}_2)=\left(\frac{2}{11},\frac{1}{66}\right)$. Therefore, for the $(n,n-1,n-1,1)$ secure exact-repair regenerating code problem, the SRK point {\em cannot} be the only corner point when $n \geq 7$.

Next, we show that both $(\bar{\alpha}_2,\bar{\beta}_2)$ and $(\bar{\alpha}_3,\bar{\beta}_3)$ are {\em optimal} tradeoff points for the $(n,n-1,n-1,1)$ secure exact-repair regenerating code problem when $n \geq 7$, so in this case the tradeoff region must have {\em multiple} corner points. 

\begin{thm}\label{thm3}
For the $(n,n-1,n-1,1)$ secure exact-repair regenerating code problem with $n \geq 7$, any achievable normalized storage-capacity repair-bandwidth pair $(\bar{\alpha},\bar{\beta}) \in \mathcal{R}_{n,n-1,n-1,1}$ must satisfy:
\begin{align}
n\bar{\alpha}+\frac{(n-1)(n-6)}{2}\bar{\beta}\geq 3.\label{eq:C3}
\end{align}
\end{thm}

Note that both
\begin{align*}
(\bar{\alpha}_2,\bar{\beta}_2)&=\left(\frac{2}{n-2},\frac{2}{(n-1)(n-2)}\right)\\
\mbox{and} \quad 
(\bar{\alpha}_3,\bar{\beta}_3)&=\left(\frac{3}{2(n-3)},\frac{3}{(n-1)(n-3)}\right)
\end{align*}
satisfy the inequality \eqref{eq:C3} with {\em equalities} and hence {\em cannot} be dominated by a single achievable tradeoff point.

Finally, we focus on the $(n,n-1,n-1,1)$ problem with $n=7$ and show that the tradeoff region has exactly {\em two} corner points at $(\bar{\alpha}_2,\bar{\beta}_2)$ and $(\bar{\alpha}_3,\bar{\beta}_3)$.

\begin{thm}\label{thm4}
For the $(7,6,6,1)$ secure exact-repair regenerating code problem, any achievable normalized storage-capacity repair-bandwidth pair $(\bar{\alpha},\bar{\beta}) \in \mathcal{R}_{7,6,6,1}$ must satisfy:
\begin{align}
\bar{\alpha} \geq \frac{3}{8}.\label{eq:C4}
\end{align}
Therefore, the tradeoff region $\mathcal{R}_{7,6,6,1}$ is given by:
\begin{align}
\mathcal{R}_{7,6,6,1}=\left\{(\bar{\alpha},\bar{\beta}): \bar{\beta} \geq \frac{1}{15},
7\bar{\alpha}+3\bar{\beta} \geq 3,
\bar{\alpha} \geq \frac{3}{8}\right\}.
\end{align}
\end{thm}

The proof of \eqref{eq:C3} and \eqref{eq:C4} can be found in Section~\ref{sec:Converse}.

\section{A New $(n,n-1,n-1,\ell)$ Code Construction}\label{sec:CodeConstr}
In this section, we provide a code construction based on the layered exact-repair regenerating codes proposed in \cite{TSAVK-IT15}, which leads to the achievability of the tradeoff points given in Theorem \ref{thm2}.

Fix a parameter $t$, and consider the following scheme. There are a total of $B={n-\ell \choose t}(t-1)$ information symbols, denoted as $\mathsf{M}$, and there are a total of $R={n \choose t}(t-1)-B$ random symbols, denoted as $\mathsf{K}$. Assume that an eavesdropper has access to the repair messages to an arbitrary set of $\ell$ nodes, the collection of which is denoted as $\mathsf{E}$. 

We first encode the $(R+B)={n\choose t}(t-1)$ symbols (in a finite field $\mathbb{F}_q$ such that $q\geq 2(R+B)$, and $q$ will be used as the basis of the entropy function), into the parity check symbols of an $(2(R+B),R+B)$ systematic MDS code. These $R+B$ symbols are broken into a total of ${n \choose t}$ parity groups, each with $(t-1)$ symbols, and each parity group is associated with a subset $\mathcal{A}$ of $[1:n]$ of cardinality $t$. 

Next, we expand each parity group by introducing one additional parity symbol which can be the simple linear sum of them, in the same finite field as earlier (or even in the binary field, assuming the original one is an extension of the binary field). These $t$ symbols are then distributed into the subset of nodes associated with this parity group, one symbol to each node. 

We need to show that the following three conditions are satisfied:
\begin{itemize}
\item[1)] Reconstruction with any $n-1$ nodes. This is trivial since in each parity group, at most one of them is in the failed node, and thus the contents of the parity group can be recovered. This also implies that
\begin{align}
\alpha_t = {n-1 \choose t-1}.
\end{align}
\item[2)] Repair with the remaining $n-1$ nodes. Assume without loss of generality that node 1 fails. Then, to repair the symbol in the parity group associated with each $\mathcal{A}$ such that $1\in \mathcal{A}$ and $| \mathcal{A}|=t$, we can send from the remaining nodes all the other symbols in this parity group. The total transmission is thus given by:
\begin{align}
(n-1)\beta_t = (t-1)\alpha_t.
\end{align}

\item[3)] Security against any eavesdropper on $\ell$ nodes. We need to show that
\begin{align}
I(\mathsf{M};\mathsf{E})=0.
\end{align}
This follows from
\begin{align*}
I(\mathsf{M};\mathsf{E})&=H(\mathsf{E})-H(\mathsf{E}|\mathsf{M})\\
&=H(\mathsf{E})-H(\mathsf{E}|\mathsf{M})+H(\mathsf{E}|\mathsf{M},\mathsf{K})\\
&=H(\mathsf{E})-I(\mathsf{E};\mathsf{K}|\mathsf{M})\\
&=H(\mathsf{E})-H(\mathsf{K}|\mathsf{M})+H(\mathsf{K}|\mathsf{M},\mathsf{E})\\
&=H(\mathsf{E})-R+H(\mathsf{K}|\mathsf{M},\mathsf{E}).
\end{align*}
All the parity groups that have symbols in the compromised nodes are completely revealed by accessing $\mathsf{E}$, and conversely all the symbols in $\mathsf{E}$ can be generated by these parity groups alone. A total of ${n \choose t}-{{n-\ell} \choose t}$ parity groups are exposed, implying that
\begin{align*}
H(\mathsf{E})\leq R. 
\end{align*}
It only remains to show that
\begin{align}
H(\mathsf{K}|\mathsf{M},\mathsf{E})=0.
\end{align}
which follows from the fact that given the eavesdropper's information and the message $\mathsf{M}$, the random symbols $\mathsf{K}$ can be completely recovered. This can be seen as follows: there are a total of $R$ symbols (after removing the simple sums in each parity group) from $\mathsf{E}$ that were original parity symbols of the $(2(R+B),R+B)$ MDS code, but any $R+B$ codeword symbols can be used to recover the original information $(\mathsf{M},\mathsf{R})$, which we indeed have together with the $B$ information symbols.
\end{itemize}
Normalizing $\alpha_t$ and $\beta_t$ by $B$ proves the achievability of the tradeoff points given in Theorem~\ref{thm2}.

\section{Proof of the Converse Results}\label{sec:Converse}
\subsection{Proof of \eqref{eq:C1} and \eqref{eq:C2}}
\begin{figure*}
\centering
\includegraphics[width=0.75\linewidth,draft=false]{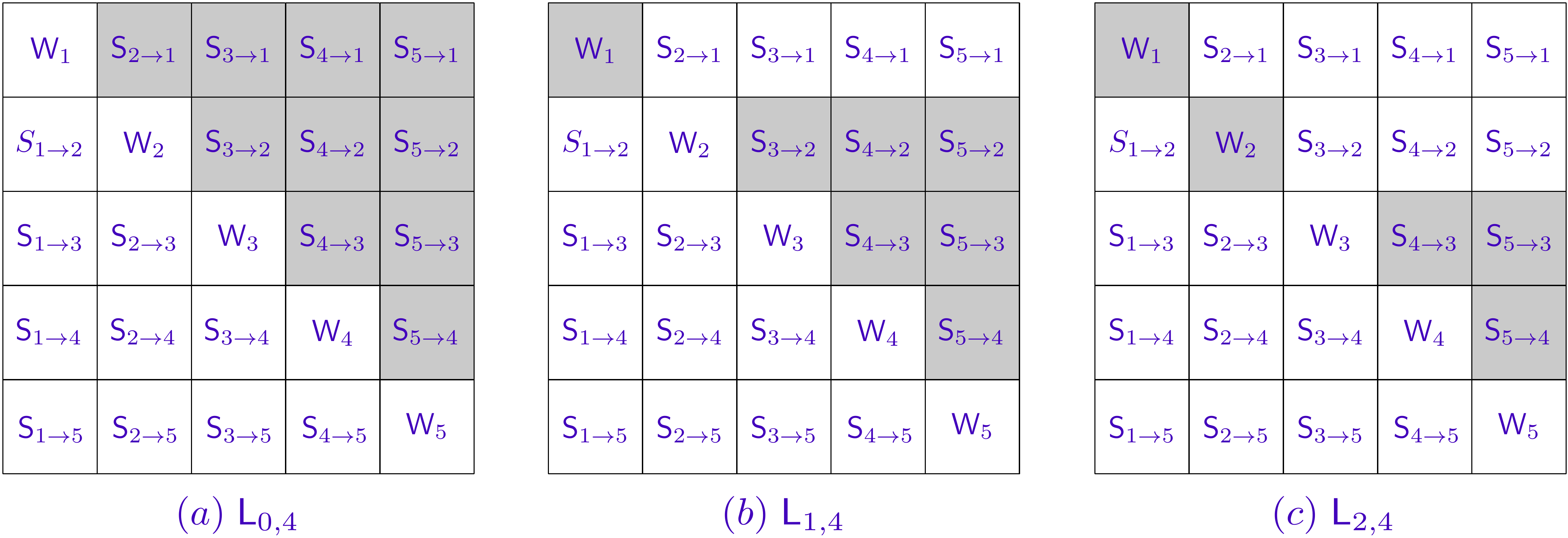}
\caption{Illustration of $\mathsf{L}_{0,4}$, $\mathsf{L}_{1,4}$ and $\mathsf{L}_{2,4}$ in the repair diagram for $n=5$.}
\label{fig2}
\end{figure*}

Let us first outline the main ingredients for proving the inequalities \eqref{eq:C1} and \eqref{eq:C2}.

\begin{itemize}
\item[1)] {\em Total number of nodes.} To prove the inequalities \eqref{eq:C1} and \eqref{eq:C2}, let us first note that these two inequalities are {\em independent} of the total number of storage nodes $n$ in the system. In our proof, we only need to consider the cases where $n=d+1$. For the cases where $n>d+1$, since any subsystem consisting of $d+1$ out of the total $n$ storage nodes must give rise to a $(d+1,k,d,\ell)$ secure exact-repair regenerating code problem. Therefore, these two inequalities as {\em outer} bounds must apply as well. When $n=d+1$, any repair group $\mathcal{B}$ of size $d$ is uniquely determined by the node $j$ to be repaired, i.e., $\mathcal{B}=[1:n]\setminus\{j\}$, and hence can be dropped from the notation $\mathsf{S}^{\mathcal{B}}_{i \rightarrow j}$ without causing any confusion. 
\item[2)] {\em Code symmetry.} Due to the built-in {\em symmetry} of the problem, to prove the inequalities \eqref{eq:C1} and \eqref{eq:C2}, we only need to consider the so-called {\em symmetrical} codes \cite{Tian-JSAC13} for which the joint entropy of any subset of random variables from 
$$\left(\mathsf{M},\mathsf{K},(\mathsf{W}_i:i\in[1:n]),(\mathsf{S}_{i \rightarrow j}: i,j\in[1:n],i\neq j)\right)$$ 
remains {\em unchanged} under any permutation over the storage-node indices. 
\item[3)] {\em Key collections of random variables.} Focusing on the symmetrical $(n=d+1,d,N,K,T,S)$ codes, the following collections of random variables play a key role in our proof:
\begin{align}
&\mathsf{W}_{\mathcal{A}} :=\left(\mathsf{W}_i:i\in \mathcal{A}\right), \quad \mathcal{A}\subseteq [1:n]\\
&\mathsf{S}_{i\rightarrow\mathcal{B}} := \left(\mathsf{S}_{i \rightarrow j}: j\in\mathcal{B}\right),\nonumber\\
&\hspace{60pt} i\in [1:n], \; \mathcal{B}\subseteq [1:n]\setminus \{i\}\\
&\mathsf{S}_{\mathcal{B}\rightarrow j} := \left(\mathsf{S}_{i \rightarrow j}: i\in\mathcal{B}\right),\nonumber\\
&\hspace{80pt} j\in[1:n],\; \mathcal{B}\subseteq [1:n]\setminus \{j\}\\
&\mathsf{S}_{\rightarrow j} := \mathsf{S}_{[1:j-1]\cup[j+1:n]\rightarrow j}, \quad j\in[1:n]\\
&\mathsf{S}_{\rightarrow \mathcal{B}} :=\left(\mathsf{S}_{\rightarrow j}:j\in\mathcal{B}\right), \quad \mathcal{B}\subseteq [1:n]\\
&\underline{\mathsf{S}}_{\rightarrow j} := \mathsf{S}_{[1:j-1]\rightarrow j}, \quad j\in[1:n]\\
&\underline{\mathsf{S}}_{\rightarrow \mathcal{B}} :=(\underline{\mathsf{S}}_{\rightarrow j}:j\in\mathcal{B}), \quad \mathcal{B}\subseteq [1:n]\\
&\overline{\mathsf{S}}_{\rightarrow j} := \mathsf{S}_{[j+1:n]\rightarrow j}, \quad j\in[1:n]\\
&\overline{\mathsf{S}}_{\rightarrow \mathcal{B}} :=(\overline{\mathsf{S}}_{\rightarrow j}:j\in\mathcal{B}), \quad \mathcal{B}\subseteq [1:n]\\
&\mathsf{L}_{t,s} :=(\mathsf{W}_{[1:t]},\overline{\mathsf{S}}_{\rightarrow[t+1:s]}),\nonumber\\
&\hspace{80pt} s\in[1:n], \; t\in[0:s]. \label{eq:DS}
\end{align}
In particular, the collection $\mathsf{L}_{t,s}$ defined in \eqref{eq:DS} was first identified in \cite{Shao-CISS16} for proving that separate encoding can achieve the MBR point for multilevel diversity coding with regeneration. As we shall see, here it also plays a key role in our proof of \eqref{eq:C1} and \eqref{eq:C2}. Fig.~\ref{fig2} illustrates the structure of $\mathsf{L}_{0,4}$, $\mathsf{L}_{1,4}$ and $\mathsf{L}_{2,4}$ in the {\em repair diagram} introduced by Duursma \cite{Duursma-P14} for $n=5$.
\end{itemize}

An important part of the proof is to understand the relations between the collections of random variables defined above, and to use them to derive the desired converse results. We have the following key lemmas, whose proof can be found in the Appendix.

\begin{lemma}\label{lemma1}
For any $(n=d+1,k,d,N,K,T,S)$ code that satisfies the node-regeneration requirement \eqref{eq:Cons3}, $(\underline{\mathsf{S}}_{\rightarrow[t+1:s]},\mathsf{W}_{[t+1:s]})$ is a function of $\mathsf{L}_{t,s}$ for any $s \in [1:n]$ and $t\in[0:s-1]$. 
\end{lemma}

\begin{lemma}\label{lemma2}
For any symmetrical $(n=d+1,k,d,N,K,T,S)$ code, we have
\begin{align}
&H(\mathsf{S}_{1 \rightarrow [2:p+1]})+H(\mathsf{L}_{t,r},\mathsf{S}_{[r+2:r+q+1] \rightarrow r+1})\nonumber\\
& \hspace{20pt} \geq H(\mathsf{S}_{1\rightarrow [2:p]})+H(\mathsf{L}_{t,r},\mathsf{S}_{[r+2:r+q+2]\rightarrow r+1})\label{eq:LLL}
\end{align}
for any $t\in[1:2]$, $r\in[2:k-1]$, $p\in[1:r-t+1]$, and $q\in[0:d-r-1]$. It follows that
\begin{align}
H(\mathsf{L}_{t,j})+T_{k,d,j}m^{-1}H(\mathsf{S}_{1\rightarrow [2:m+1]}) & \geq H(\mathsf{L}_{t,k})\label{eq:P3}
\end{align}
for any $t\in[1:2]$, $j\in[2:k]$, and $m\in[1:j-t+1]$.
\end{lemma}

\begin{lemma}\label{lemma3}
For any symmetrical $(n=d+1,k,d,N,K,T,S)$ code that satisfies the node-regeneration requirement \eqref{eq:Cons3}, we have
\begin{align}
&\frac{d-t}{n-j}H(\mathsf{L}_{1,j},\mathsf{S}_{j\rightarrow 1})+H(\mathsf{L}_{1,t})\nonumber\\
&\hspace{20pt} \geq \frac{d-t}{n-j}H(\mathsf{L}_{1,j-1},\mathsf{S}_{j\rightarrow 1})+H(\mathsf{L}_{1,t+1})
\label{eq:exchange3}
\end{align}
for any $j\in[2:k-1]$ and $t\in[j:k-1]$. It follows that
\begin{align}
&H(\mathsf{L}_{1,j},\mathsf{S}_{j\rightarrow 1})+(n-j)T_{k,d,m}^{-1}H(\mathsf{L}_{1,m})\nonumber\\
& \hspace{20pt} \geq H(\mathsf{L}_{1,j-1},\mathsf{S}_{j\rightarrow 1})+(n-j)T_{k,d,m}^{-1}H(\mathsf{L}_{1,k})\label{eq:P4}
\end{align}
for any $j\in[2:k-1]$ and $m\in[j:k-1]$.
\end{lemma}

The inequality \eqref{eq:C1} can now be proved as follows:
\begin{align*}
B&=H(\mathsf{M})\\
& \stackrel{(a)}{=} H(\mathsf{M}|\mathsf{S}_{\rightarrow[1:\ell]})\\
& \stackrel{(b)}{=} H(\mathsf{M}|\mathsf{S}_{\rightarrow[1:\ell]},\mathsf{W}_{[1:\ell]})\\
&\leq H(\mathsf{M},\overline{\mathsf{S}}_{\rightarrow[\ell+1:k]}|\mathsf{S}_{\rightarrow[1:\ell]},\mathsf{W}_{[1:\ell]})\\
&= H(\mathsf{M}|\mathsf{S}_{\rightarrow[1:\ell]},\mathsf{W}_{[1:\ell]},\overline{\mathsf{S}}_{\rightarrow[\ell+1:k]})\\
&\hspace{20pt}+H(\bar{\mathsf{S}}_{\rightarrow[\ell+1:k]}|\mathsf{S}_{\rightarrow[1:\ell]},\mathsf{W}_{[1:\ell]})\\
&= H(\mathsf{M}|\mathsf{S}_{\rightarrow[1:\ell]},\mathsf{L}_{\ell,k})+H(\bar{\mathsf{S}}_{\rightarrow[\ell+1:k]}|\mathsf{S}_{\rightarrow[1:\ell]},\mathsf{W}_{[1:\ell]})\\
& \stackrel{(c)}{=} H(\mathsf{M}|\mathsf{S}_{\rightarrow[1:\ell]},\mathsf{L}_{\ell,k},\mathsf{W}_{[\ell+1:k]})\\
&\hspace{20pt}+H(\bar{\mathsf{S}}_{\rightarrow[\ell+1:k]}|\mathsf{S}_{\rightarrow[1:\ell]},\mathsf{W}_{[1:\ell]})\\
&= H(\mathsf{M}|\mathsf{S}_{\rightarrow[1:\ell]},\overline{\mathsf{S}}_{\rightarrow[\ell+1:k]},\mathsf{W}_{[1:k]})\\
&\hspace{20pt}+H(\bar{\mathsf{S}}_{\rightarrow[\ell+1:k]}|\mathsf{S}_{\rightarrow[1:\ell]},\mathsf{W}_{[1:\ell]})\\
& \stackrel{(d)}{=} H(\bar{\mathsf{S}}_{\rightarrow[\ell+1:k]}|\mathsf{S}_{\rightarrow[1:\ell]},\mathsf{W}_{[1:\ell]})\\
&\leq H(\bar{\mathsf{S}}_{\rightarrow[\ell+1:k]})\\
& \stackrel{(d)}{\leq} T_{k,d,l}\beta
\end{align*}
where $(a)$ follows from the repair-secrecy constraint \eqref{eq:Cons4}; $(b)$ follows from the fact that $\mathsf{W}_{[1:\ell]}$ is a function of $\mathsf{S}_{\rightarrow[1:\ell]}$ due to the node-regeneration constraint \eqref{eq:Cons3}; $(c)$ follows from the fact that $\mathsf{W}_{[\ell+1:k]})$ is a function of $\mathsf{L}_{\ell,k}$ by Lemma~\ref{lemma1}; $(d)$ follows from the fact that $H(\mathsf{M}|\mathsf{S}_{\rightarrow[1:\ell]},\overline{\mathsf{S}}_{\rightarrow[\ell+1:k]},\mathsf{W}_{[1:k]})=0$ due to the message-recovery constraint \eqref{eq:Cons2}; and $(e)$ follows from the bandwidth constraint on the repair messages. Normalizing both sides by $B$ completes the proof of \eqref{eq:C1}.

To prove the inequality \eqref{eq:C2}, we shall consider the cases where $T_{k,d,\ell} \leq d$ and $d \leq T_{k,d,\ell} \leq d+\sqrt{d\ell}$ separately. 

Case 1: $T_{k,d,\ell} \leq d$. In this case, we have
\begin{align}
&T_{k,d,\ell}\alpha+dH(\mathsf{S}_{\rightarrow [1:\ell]})\nonumber\\
&=T_{k,d,\ell}\left(\alpha+H(\mathsf{S}_{\rightarrow [1:\ell]})\right)+\left(d-T_{k,d,\ell}\right)H(\mathsf{S}_{\rightarrow [1:\ell]})\nonumber\\
&\stackrel{(a)}{=}T_{k,d,\ell}\left(\alpha+H(\mathsf{S}_{\rightarrow [2:\ell+1]}\right)+\left(d-T_{k,d,\ell}\right)H(\mathsf{S}_{\rightarrow [1:\ell]})\nonumber\\
&\stackrel{(b)}{\geq} T_{k,d,\ell}\left(H(\mathsf{W}_1)+H(\mathsf{S}_{\rightarrow [2:\ell+1]})\right)+\left(d-T_{k,d,\ell}\right)H(\mathsf{S}_{\rightarrow [1:\ell]})\nonumber\\
&\stackrel{(c)}{=} T_{k,d,\ell}\left(H(\mathsf{W}_1,\mathsf{S}_{1\rightarrow [2:\ell+1]})+H(\mathsf{S}_{\rightarrow [2:\ell+1]})\right)\nonumber\\
& \hspace{13pt} +\left(d-T_{k,d,\ell}\right)H(\mathsf{S}_{\rightarrow [1:\ell]})\nonumber\\
&\stackrel{(d)}{\geq} T_{k,d,\ell}\left(H(\mathsf{W}_1,\mathsf{S}_{\rightarrow [2:\ell+1]})+H(\mathsf{S}_{1\rightarrow [2:\ell+1]})\right)\nonumber\\
& \hspace{13pt} +\left(d-T_{k,d,\ell}\right)H(\mathsf{S}_{\rightarrow [1:\ell]})\nonumber\\
&\stackrel{(e)}{=} T_{k,d,\ell}\left(H(\mathsf{W}_1,\mathsf{S}_{\rightarrow [2:\ell+1]})+T_{k,d,\ell+1}\ell^{-1}H(\mathsf{S}_{1\rightarrow [2:\ell+1]})\right)\nonumber\\
& \hspace{13pt} +\left(d-T_{k,d,\ell}\right)\left(H(\mathsf{S}_{\rightarrow[1:\ell]})+T_{k,d,\ell}\ell^{-1}H(\mathsf{S}_{1\rightarrow [2:\ell+1]})\right)\nonumber\\
&\stackrel{(f)}{\geq} T_{k,d,\ell}\left(H(\mathsf{L}_{1,\ell+1})+T_{k,d,\ell+1}\ell^{-1}H(\mathsf{S}_{1\rightarrow [2:\ell+1]})\right)\nonumber\\
& \hspace{13pt} +\left(d-T_{k,d,\ell}\right)\left(H(\mathsf{L}_{1,\ell})+T_{k,d,\ell}\ell^{-1}H(\mathsf{S}_{1\rightarrow [2:\ell+1]})\right)\label{eq:ZZ1}
\end{align}
where $(a)$ follows from the fact that
\begin{align}
H(\mathsf{S}_{\rightarrow [1:\ell]})=H(\mathsf{S}_{\rightarrow [2:\ell+1]})\label{eq:ZZ-Sym}
\end{align}
due to the symmetrical code that we consider; $(b)$ is due to the storage-capacity constraint $H(\mathsf{W}_1) \leq \alpha$; $(c)$ is due to the fact that $\mathsf{S}_{1\rightarrow [2:\ell+1]}$ is a function of $\mathsf{W}_1$; $(d)$ follows from the fact that 
\begin{align}
H(\mathsf{W}_1,&\mathsf{S}_{1\rightarrow[2:\ell+1]})+H(\mathsf{S}_{\rightarrow [2:\ell+1]})\nonumber\\
& \geq H(\mathsf{W}_1,\mathsf{S}_{\rightarrow [2:\ell+1]})+H(\mathsf{S}_{1\rightarrow [2:\ell+1]})\label{eq:ZZ-Sub}
\end{align}
due to the submodularity of the entropy function; $(e)$ follows from the fact that 
$$T_{k,d,\ell+1}\ell^{-1}+\left(d-T_{k,d,\ell}\right)\ell^{-1}=1;$$ 
and $(f)$ follows from the facts that 
\begin{align}
& H(\mathsf{W}_1,\mathsf{S}_{\rightarrow[2:\ell+1]}) \geq H(\mathsf{W}_1,\overline{\mathsf{S}}_{\rightarrow[2:\ell+1]})=H(\mathsf{L}_{1,\ell+1})\label{eq:ZZ-P1}\\
& H(\mathsf{S}_{\rightarrow[1:\ell]})=H(\mathsf{W}_1,\mathsf{S}_{\rightarrow[1:\ell]})\nonumber\\
& \hspace{80pt} \geq H(\mathsf{W}_1,\overline{\mathsf{S}}_{\rightarrow[2:\ell]})=H(\mathsf{L}_{1,\ell}).\label{eq:ZZ-P2}
\end{align}

Applying \eqref{eq:P3} with $(t,j,m)=(1,\ell+1,\ell)$ and $(t,j,m)=(1,\ell,\ell)$, respectively gives:
\begin{align}
H(\mathsf{L}_{1,\ell+1})+T_{k,d,\ell+1}\ell^{-1}H(\mathsf{S}_{1\rightarrow [2:\ell+1]}) & \geq H(\mathsf{L}_{1,k})\label{eq:ZZ2}\\
H(\mathsf{L}_{1,\ell})+T_{k,d,\ell+1}\ell^{-1}H(\mathsf{S}_{1\rightarrow [2:\ell+1]}) & \geq H(\mathsf{L}_{1,k}).\label{eq:ZZ3}
\end{align}
Substituting \eqref{eq:ZZ2} and \eqref{eq:ZZ3} into \eqref{eq:ZZ1} gives:
\begin{align*}
&T_{k,d,\ell}\alpha+dH(\mathsf{S}_{\rightarrow[1:\ell]}) \geq dH(\mathsf{L}_{1,k})\\
&\stackrel{(a)}{=}dH(\mathsf{L}_{1,k},\mathsf{W}_{[2:k]},\underline{\mathsf{S}}_{\rightarrow [2:k]})=dH(\mathsf{W}_{[1:k]},\mathsf{S}_{\rightarrow[2:k]})\\
&\stackrel{(b)}{=}dH(\mathsf{W}_{[1:k]},\mathsf{M},\mathsf{S}_{\rightarrow[2:k]})\geq dH(\mathsf{M},\mathsf{S}_{\rightarrow [2:\ell+1]})\\
&=dH(\mathsf{S}_{\rightarrow [2:\ell+1]})+dH(\mathsf{M}|\mathsf{S}_{\rightarrow [2:\ell+1]})\\
&\stackrel{(c)}{=}dH(\mathsf{S}_{\rightarrow [2:\ell+1]})+dH(\mathsf{M}) \stackrel{(d)}{=}dH(\mathsf{S}_{\rightarrow [1:\ell]})+dB
\end{align*}
where $(a)$ follows from the fact that $(\mathsf{W}_{[2:k]},\underline{\mathsf{S}}_{\rightarrow [2:k]})$ is a function of $\mathsf{L}_{1,k}$ by Lemma~\ref{lemma1}; $(b)$ follows from the fact that $\mathsf{M}$ is a function of $\mathsf{W}_{[1:k]}$ due to the message-recover constraint \eqref{eq:Cons2}; $(c)$ follows from the repair-secrecy constraint \eqref{eq:Cons4}; and $(d)$ follows again from \eqref{eq:ZZ-Sym} due to the symmetrical code that we consider. Canceling $dH(\mathsf{S}_{\rightarrow [1:\ell]})$ from both sides of the inequality completes the proof of \eqref{eq:C2} for the cases where $T_{k,d,\ell} \leq d$.

Case 2: $d \leq T_{k,d,\ell} \leq d+\sqrt{d\ell}$. Note that if $k=\ell+1$, we have $T_{k,d,\ell}=d-\ell< d$. Therefore, in this case we must have $k \geq \ell+2\geq 3$. In addition, let 
\begin{align*}
q:=1+(d-\ell)T_{k,d,\ell+1}^{-1}
\end{align*}
and we have
\begin{align*}
d&-\left(T_{k,d,\ell}-d\right)q\\
&=T_{k,d,\ell+1}^{-1}\left(-T_{k,d,\ell}^2+2dT_{k,d,\ell}-d(d-\ell)\right) \geq 0.
\end{align*}
It follows that
\begin{align}
&T_{k,d,\ell}\alpha+dH(\mathsf{S}_{\rightarrow[1:\ell]})\nonumber\\
&=d\left(\alpha+H(\mathsf{S}_{\rightarrow [1:\ell]})\right)+\left(T_{k,d,\ell}-d\right)\alpha\nonumber\\
&\stackrel{(a)}{=}d\left(\alpha+H(\mathsf{S}_{\rightarrow [2:\ell+1]})\right)+\left(T_{k,d,\ell}-d\right)\alpha\nonumber\\
&\stackrel{(b)}{\geq} d\left(H(\mathsf{W}_1)+H(\mathsf{S}_{\rightarrow [2:\ell+1]})\right)+\left(T_{k,d,\ell}-d\right)\alpha\nonumber\\
&\stackrel{(c)}{=} d\left(H(\mathsf{W}_1,\mathsf{S}_{1\rightarrow [2:\ell+1]})+H(\mathsf{S}_{\rightarrow [2:\ell+1]})\right)+\left(T_{k,d,\ell}-d\right)\alpha\nonumber\\
&\stackrel{(d)}{\geq} d\left(H(\mathsf{W}_1,\mathsf{S}_{\rightarrow [2:\ell+1]})+H(\mathsf{S}_{1\rightarrow [2:\ell+1]})\right)+\left(T_{k,d,\ell}-d\right)\alpha\nonumber\\
&\stackrel{(e)}{\geq} d\left(H(\mathsf{L}_{1,\ell+1})+H(\mathsf{S}_{1\rightarrow [2:\ell+1]})\right)+\left(T_{k,d,\ell}-d\right)\alpha\nonumber\\
&=\left(d-\left(T_{k,d,\ell}-d\right)q\right)H(\mathsf{L}_{1,\ell+1})\nonumber\\
&\hspace{13pt}+\left(T_{k,d,\ell}-d\right)\left(qH(\mathsf{L}_{1,\ell+1})+\alpha\right)+dH(\mathsf{S}_{1\rightarrow [2:\ell+1]})\label{eq:WW1}
\end{align}
where $(a)$ follows from \eqref{eq:ZZ-Sym} due to the symmetrical code that we consider; $(b)$ is due to the storage-capacity constraint $H(\mathsf{W}_1) \leq \alpha$; $(c)$ is due to the fact that $\mathsf{S}_{1\rightarrow [2:\ell+1]}$ is a function of $\mathsf{W}_1$; $(d)$ follows from \eqref{eq:ZZ-Sub} due to the submodularity of the entropy function; and $(e)$ follows from \eqref{eq:ZZ-P1}.

The first term on the right-hand side of \eqref{eq:WW1} can be further bounded from below by the fact that $\mathsf{L}_{2,\ell+1}$ is a function of $\mathsf{L}_{1,\ell+1}$ by Lemma~\ref{lemma1}, so we have
\begin{align}
H(\mathsf{L}_{1,\ell+1}) \geq H(\mathsf{L}_{2,\ell+1}).\label{eq:WW2}
\end{align}
To bound from below the second term on the right-hand side of \eqref{eq:WW1}, note that
\begin{align}
&qH(\mathsf{L}_{1,\ell+1})+\alpha\nonumber\\
& =H(\mathsf{L}_{1,\ell+1})+(d-\ell)T_{k,d,\ell+1}^{-1}H(\mathsf{L}_{1,\ell+1})+\alpha\nonumber\\
& \stackrel{(a)}{=}H(\mathsf{L}_{1,\ell+1},\mathsf{S}_{\ell+1\rightarrow1})+(d-\ell)T_{k,d,\ell+1}^{-1}H(\mathsf{L}_{1,\ell+1})+\alpha\nonumber\\
& \stackrel{(b)}{\geq} H(\mathsf{L}_{1,\ell},\mathsf{S}_{\ell+1\rightarrow 1})+(d-\ell)T_{k,d,\ell+1}^{-1}H(\mathsf{L}_{1,k})+\alpha\nonumber\\
& \stackrel{(c)}{\geq} H(\mathsf{L}_{1,\ell},\mathsf{S}_{\ell+1\rightarrow 1})+(d-\ell)T_{k,d,\ell+1}^{-1}H(\mathsf{L}_{1,k})+H(\mathsf{W}_{\ell+1})\nonumber\\
& \stackrel{(d)}{=} H(\mathsf{L}_{1,\ell},\mathsf{S}_{\ell+1\rightarrow 1})+H(\mathsf{W}_{\ell+1},\mathsf{S}_{\ell+1\rightarrow [1:\ell]})+\nonumber\\
& \hspace{13pt} (d-\ell)T_{k,d,\ell+1}^{-1}H(\mathsf{L}_{1,k})\nonumber\\
& \stackrel{(e)}{\geq} H(\mathsf{L}_{1,\ell},\mathsf{W}_{\ell+1},\mathsf{S}_{\ell+1\rightarrow 1})+H(\mathsf{S}_{\ell+1\rightarrow [1:\ell]})+\nonumber\\
& \hspace{13pt} (d-\ell)T_{d,\ell+1,k}^{-1}H(\mathsf{L}_{1,k})\nonumber\\
& \stackrel{(f)}{=} H(\mathsf{L}_{1,\ell},\mathsf{W}_{\ell+1})+H(\mathsf{S}_{\ell+1\rightarrow [1:\ell]})+(d-l)T_{k,d,l+1}^{-1}H(\mathsf{L}_{1,k})\nonumber\\
& \stackrel{(g)}{=} H(\mathsf{L}_{2,\ell+1})+H(\mathsf{S}_{1\rightarrow [2:\ell+1]})+(d-\ell)T_{k,d,\ell+1}^{-1}H(\mathsf{L}_{1,k})\nonumber\\
& \stackrel{(h)}{\geq} H(\mathsf{L}_{2,\ell+1})+H(\mathsf{S}_{1\rightarrow [2:\ell+1]})+(d-\ell)T_{k,d,\ell+1}^{-1}H(\mathsf{L}_{2,k})\label{eq:WW3}
\end{align}
where $(a)$ follows from the fact that $\mathsf{S}_{\ell+1\rightarrow1}$ is a function of $\mathsf{W}_{\ell+1}$, which is in turn a function of $\mathsf{L}_{1,\ell+1}$ by Lemma~\ref{lemma1}; (b) follows from \eqref{eq:P4} with $(j,m)=(\ell+1,\ell+1)$; $(c)$ is due to the storage-capacity constraint $H(\mathsf{W}_{\ell+1}) \leq \alpha$; $(d)$ follows from the fact that $\mathsf{S}_{\ell+1\rightarrow1}$ is a function of $\mathsf{W}_{\ell+1}$; $(e)$ follows from the fact that
\begin{align}
H(\mathsf{L}_{1,\ell},&\mathsf{S}_{\ell+1\rightarrow 1})+H(\mathsf{W}_{\ell+1},\mathsf{S}_{\ell+1\rightarrow [1:\ell]})\nonumber\\
& \geq H(\mathsf{L}_{1,\ell},\mathsf{W}_{\ell+1},\mathsf{S}_{\ell+1\rightarrow 1})+H(\mathsf{S}_{\ell+1\rightarrow [1:\ell]})
\end{align}
due to the submodularity of the entropy function; $(f)$ follows yet again from the fact that $\mathsf{S}_{\ell+1\rightarrow 1}$ is a function of $\mathsf{W}_{\ell+1}$; $(g)$ follows from the facts that
\begin{align}
H(\mathsf{L}_{1,\ell},\mathsf{W}_{\ell+1}) =H(\mathsf{L}_{2,\ell+1})\\
H(\mathsf{S}_{\ell+1\rightarrow [1:\ell]}) =H(\mathsf{S}_{1\rightarrow [2:\ell+1]})
\end{align}
due to the symmetrical code that we consider; and $(h)$ follows from the fact that $\mathsf{L}_{2,k}$ is a function of $\mathsf{L}_{1,k}$ by Lemma~\ref{lemma1}, so we have
\begin{align}
H(\mathsf{L}_{1,k}) \geq H(\mathsf{L}_{2,k}).
\end{align}

Substituting \eqref{eq:WW2} and \eqref{eq:WW3} into \eqref{eq:WW1} gives:
\begin{align*}
&T_{k,d,\ell}\alpha+dH(\mathsf{S}_{\rightarrow [1:\ell]})\nonumber\\
&\geq \left(d-(T_{k,d,\ell}-d)(d-\ell)T_{k,d,\ell+1}^{-1}\right)H(\mathsf{L}_{2,\ell+1})+\nonumber\\
&\hspace{13pt} T_{k,d,\ell}H(\mathsf{S}_{1\rightarrow [2:\ell+1]})+\left(T_{k,d,\ell}-d\right)(d-\ell)T_{k,d,\ell+1}^{-1}H(\mathsf{L}_{2,k})\nonumber\\
&\stackrel{(a)}{=} T_{k,d,\ell}T_{k,d,\ell+1}^{-1}\ell\left(H(\mathsf{L}_{2,\ell+1})+T_{k,d,\ell+1}\ell^{-1}H(\mathsf{S}_{1\rightarrow [2:\ell+1]}\right)+\nonumber\\
&\hspace{13pt} \left(T_{k,d,\ell}-d\right)(d-\ell)T_{k,d,\ell+1}^{-1}H(\mathsf{L}_{2,k})\nonumber\\
& \stackrel{(b)}{\geq} \left(T_{k,d,\ell}T_{k,d,\ell+1}^{-1}\ell+(T_{k,d,\ell}-d)(d-\ell)T_{k,d,\ell+1}^{-1}\right)H(\mathsf{L}_{2,k})\nonumber\\
& \stackrel{(c)}{=} dH(\mathsf{L}_{2,k})  \stackrel{(d)}{=}dH(\mathsf{L}_{2,k},\mathsf{W}_{[3:k]},\underline{\mathsf{S}}_{\rightarrow[3:k]})\\
& =dH(\mathsf{W}_{[1:k]},\mathsf{S}_{\rightarrow [3:k]})\stackrel{(e)}{=} dH(\mathsf{W}_{[1:k]},\mathsf{M},\mathsf{S}_{\rightarrow [3:k]})\\
& \geq dH(\mathsf{M},\mathsf{S}_{\rightarrow [3:\ell+2]})=dH(\mathsf{S}_{\rightarrow [3:\ell+2]})+dH(\mathsf{M}|\mathsf{S}_{\rightarrow [3:\ell+2]})\\
& \stackrel{(f)}{=}dH(\mathsf{S}_{\rightarrow [3:\ell+2]})+dH(\mathsf{M})\stackrel{(g)}{=}dH(\mathsf{S}_{\rightarrow [1:\ell]})+dB
\end{align*}
where $(a)$ and $(c)$ follow from the fact that
\begin{align*}
d-(T_{k,d,\ell}-d)(d-\ell)T_{k,d,\ell+1}^{-1}&=T_{k,d,\ell}T_{k,d,\ell+1}^{-1}\ell;
\end{align*}
$(b)$ follows from \eqref{eq:P3} with $(t,j,m)=(2,\ell+1,l)$; $(d)$ follows from the fact that $(\mathsf{W}_{[3:k]},\underline{\mathsf{S}}_{\rightarrow[3:k]})$ is a function of $\mathsf{L}_{2,k}$ by Lemma~\ref{lemma1}; $(e)$ follows from the fact that $\mathsf{M}$ is a function of $\mathsf{W}_{[1:k]}$ due to the message-recover constraint \eqref{eq:Cons2}; $(f)$ is due to the secrecy constraint \eqref{eq:Cons4}; and $(g)$ follows from the fact that
\begin{align}
H(\mathsf{S}_{\rightarrow [3:\ell+2]})=H(\mathsf{S}_{\rightarrow [1:\ell]})
\end{align}
due to the symmetrical code that we consider. Canceling $dH(\mathsf{S}_{\rightarrow [1:\ell]})$ from both sides of the inequality completes the proof of \eqref{eq:C2} for $d \leq T_{k,d,\ell} \leq d+\sqrt{d\ell}$.

\subsection{Proof of \eqref{eq:C3}}
Assume that $k=d=n-1$ and $\ell=1$. As before, we shall also assume without loss of generality that the codes that we consider here are symmetrical ones. 

Let us first show that for any $i\in[1:n-1]$, we have
\begin{align}
H(\mathsf{S}_{\rightarrow 1}) \geq H(\overline{\mathsf{S}}_{\rightarrow i}|\mathsf{W}_{[1:i-1]})+H(\underline{\mathsf{S}}_{\rightarrow i})\label{eq:Ph1}
\end{align}
which can be seen as follows:
\begin{align*}
H&(\mathsf{W}_{[1:i-1]})+H(\mathsf{S}_{\rightarrow 1})\\
& \stackrel{(a)}{=} H(\mathsf{W}_{[1:i-1]})+H(\mathsf{S}_{\rightarrow i})\\
& \stackrel{(b)}{=} H(\mathsf{W}_{[1:i-1]},\underline{\mathsf{S}}_{\rightarrow i})+H(\underline{\mathsf{S}}_{\rightarrow i},\overline{\mathsf{S}}_{\rightarrow i})\\
& \stackrel{(c)}{\geq} H(\mathsf{W}_{[1:i-1]},\underline{\mathsf{S}}_{\rightarrow i},\overline{\mathsf{S}}_{\rightarrow i})+H(\underline{\mathsf{S}}_{\rightarrow i})\\
& \geq H(\mathsf{W}_{[1:i-1]},\overline{\mathsf{S}}_{\rightarrow i})+H(\underline{\mathsf{S}}_{\rightarrow i})\\
& \geq H(\mathsf{W}_{[1:i-1]})+H(\overline{\mathsf{S}}_{\rightarrow i}|\mathsf{W}_{[1:i-1]})+H(\underline{\mathsf{S}}_{\rightarrow i})
\end{align*}
where $(a)$ follows from the fact that $H(\mathsf{S}_{\rightarrow 1})=H(\mathsf{S}_{\rightarrow i})$ due to the symmetrical code that we consider; $(b)$ is due to the fact that $\underline{\mathsf{S}}_{\rightarrow i}$ is a function of $\mathsf{W}_{[1:i-1]}$; and $(c)$ is due to the submodularity of the entropy function. Canceling $H(\mathsf{W}_{[1:i-1]})$ from both sides completes the proof of \eqref{eq:Ph1}.

Setting $i=2,3$ in \eqref{eq:Ph1} and by the symmetrical code that we consider, we have
\begin{align}
H(\mathsf{S}_{\rightarrow 1}) & \geq H(\overline{\mathsf{S}}_{\rightarrow 2}|\mathsf{W}_{1})+H(\underline{\mathsf{S}}_{\rightarrow 2})\nonumber\\
& = H(\overline{\mathsf{S}}_{\rightarrow 2}|\mathsf{W}_{1})+H(\mathsf{S}_{1 \rightarrow 2})\nonumber\\
& = H(\overline{\mathsf{S}}_{\rightarrow 2}|\mathsf{W}_{1})+H(\mathsf{S}_{n \rightarrow n-1})\nonumber\\
& = H(\overline{\mathsf{S}}_{\rightarrow 2}|\mathsf{W}_{1})+H(\overline{\mathsf{S}}_{\rightarrow n-1})\nonumber\\
& \geq H(\overline{\mathsf{S}}_{\rightarrow 2}|\mathsf{W}_{1})+H(\overline{\mathsf{S}}_{\rightarrow n-1}|\mathsf{W}_{[1:n-2]})\label{eq:Ph2}
\end{align}
and
\begin{align}
H(\mathsf{S}_{\rightarrow 1}) & \geq H(\overline{\mathsf{S}}_{\rightarrow 3}|\mathsf{W}_{[1:2]})+H(\underline{\mathsf{S}}_{\rightarrow 3})\nonumber\\
& = H(\overline{\mathsf{S}}_{\rightarrow 3}|\mathsf{W}_{[1:2]})+H(\mathsf{S}_{[1:2] \rightarrow 3})\nonumber\\
& = H(\overline{\mathsf{S}}_{\rightarrow 3}|\mathsf{W}_{[1:2]})+H(\mathsf{S}_{[n-1:n] \rightarrow n-2})\nonumber\\
& = H(\overline{\mathsf{S}}_{\rightarrow 3}|\mathsf{W}_{[1:2]})+H(\overline{\mathsf{S}}_{\rightarrow n-2})\nonumber\\
& \geq H(\overline{\mathsf{S}}_{\rightarrow 3}|\mathsf{W}_{[1:2]})+H(\overline{\mathsf{S}}_{\rightarrow n-2}|\mathsf{W}_{[1:n-3]}).\label{eq:Ph3}
\end{align}
Adding \eqref{eq:Ph2} and \eqref{eq:Ph3} gives:
\begin{align}
3&H(\mathsf{S}_{\rightarrow 1})\nonumber\\
& \geq H(\mathsf{S}_{\rightarrow 1})+H(\overline{\mathsf{S}}_{\rightarrow 2}|\mathsf{W}_{1})+H(\overline{\mathsf{S}}_{\rightarrow 3}|\mathsf{W}_{[1:2]})+\nonumber\\
&\hspace{13pt}H(\overline{\mathsf{S}}_{\rightarrow n-2}|\mathsf{W}_{[1:n-3]})+H(\overline{\mathsf{S}}_{\rightarrow n-1}|\mathsf{W}_{[1:n-2]}).\label{eq:Ph4}
\end{align}

Furthermore, by the repair-bandwidth constraint, we have
\begin{align}
\frac{(n-1)(n-6)}{2}\beta & = \sum_{i=3}^{n-4}i\beta \geq \sum_{i=3}^{n-4}H(\overline{\mathsf{S}}_{\rightarrow n-i})\nonumber\\
& \geq \sum_{i=3}^{n-4}H(\overline{\mathsf{S}}_{\rightarrow n-i}|\mathsf{W}_{[1:n-i-1]}).\label{eq:Ph5}
\end{align}
Adding \eqref{eq:Ph4} and \eqref{eq:Ph5} gives:
\begin{align}
&\frac{(n-1)(n-6)}{2}\beta+3H(\mathsf{S}_{\rightarrow 1})\nonumber\\
& \geq \sum_{i=1}^{n-1}H(\overline{\mathsf{S}}_{\rightarrow n-i}|\mathsf{W}_{[1:n-i-1]}) = \sum_{i=1}^{n-1}H(\overline{\mathsf{S}}_{\rightarrow i}|\mathsf{W}_{[1:i-1]})\label{eq:Ph6}
\end{align}
where the last equality follows from the change of variable $i \rightarrow n-i$. 

To proceed, we shall need the following lemma, whose proof can be found in the Appendix.

\begin{lemma}\label{lemma4}
For any symmetrical $(n,k=n-1,d=n-1,N,K,T,S)$ code that satisfies the node-regeneration requirement \eqref{eq:Cons3} and the repair-secrecy constraint \eqref{eq:Cons4} with $\ell=1$, we have
\begin{align}
&\sum_{i=1}^{n-1}H(\overline{\mathsf{S}}_{\rightarrow i}|\mathsf{W}_{[1:i-1]})+n\alpha \geq 3H(\mathsf{S}_{\rightarrow 1})+3B.\label{eq:lemma4}
\end{align}
\end{lemma}
Adding \eqref{eq:Ph6} and \eqref{eq:lemma4} gives:
\begin{align*}
\frac{(n-1)(n-6)}{2}\beta+n\alpha+3H(\mathsf{S}_{\rightarrow 1}) \geq 3H(\mathsf{S}_{\rightarrow 1})+3B.
\end{align*}
Canceling $3H(\mathsf{S}_{\rightarrow 1})$ from both sides and normalizing the remaining terms by $B$ complete the proof of \eqref{eq:C3}.

\subsection{Proof of \eqref{eq:C4}}
Let us first show that
\begin{align}
3H(\mathsf{S}_{\rightarrow 1}) &\geq \sum_{i=2}^{6}H(\overline{\mathsf{S}}_{\rightarrow i}|\mathsf{W}_{[1:i-1]})+H(\underline{\mathsf{S}}_{\rightarrow 4})\label{eq:BBQ}
\end{align}
which can be seen as follows. 

First note that
\begin{align}
H(\mathsf{S}_{\rightarrow 1}) & \stackrel{(a)}{\geq} H(\overline{\mathsf{S}}_{\rightarrow 2}|\mathsf{W}_1)+H(\mathsf{S}_{1 \rightarrow 2})\nonumber\\
& \stackrel{(b)}{=} H(\overline{\mathsf{S}}_{\rightarrow 2}|\mathsf{W}_1)+H(\mathsf{S}_{7 \rightarrow 6})\nonumber\\
& = H(\overline{\mathsf{S}}_{\rightarrow 2}|\mathsf{W}_1)+H(\overline{\mathsf{S}}_{\rightarrow 6})\nonumber\\
& \geq H(\overline{\mathsf{S}}_{\rightarrow 2}|\mathsf{W}_1)+H(\overline{\mathsf{S}}_{\rightarrow 6}|\mathsf{W}_{[1:5]})
\label{eq:BBQ1}
\end{align}
where $(a)$ follows from \eqref{eq:Ph1} with $n=7$ and $i=2$; and $(b)$ follows from the fact that $H(\mathsf{S}_{1\rightarrow 2})=H(\mathsf{S}_{7\rightarrow 6})$ due to the symmetrical code that we consider. Next, we have
\begin{align}
H(\mathsf{S}_{\rightarrow 1}) & \stackrel{(a)}{\geq} H(\overline{\mathsf{S}}_{\rightarrow 3}|\mathsf{W}_{[1:2]})+H(\underline{\mathsf{S}}_{\rightarrow 3})\nonumber\\
&= H(\overline{\mathsf{S}}_{\rightarrow 3}|\mathsf{W}_{[1:2]})+H(\mathsf{S}_{[1:2]\rightarrow 3})\nonumber\\
&\stackrel{(b)}{=} H(\overline{\mathsf{S}}_{\rightarrow 3}|\mathsf{W}_{[1:2]})+H(\mathsf{S}_{[6:7]\rightarrow 5})\nonumber\\
&= H(\overline{\mathsf{S}}_{\rightarrow 3}|\mathsf{W}_{[1:2]})+H(\overline{\mathsf{S}}_{\rightarrow 5})\nonumber\\
&\geq H(\overline{\mathsf{S}}_{\rightarrow 3}|\mathsf{W}_{[1:2]})+H(\overline{\mathsf{S}}_{\rightarrow 5}|\mathsf{W}_{[1:4]})\label{eq:BBQ2}
\end{align}
where $(a)$ follows from \eqref{eq:Ph1} with $n=7$ and $i=3$; and $(b)$ follows from the fact that $H(\mathsf{S}_{[1:2]\rightarrow 3})=H(\mathsf{S}_{[6:7]\rightarrow 5})$ due to the symmetrical code that we consider. Finally, setting $n=7$ and $i=4$ in \eqref{eq:Ph1} gives
\begin{align}
H(\mathsf{S}_{\rightarrow 1}) \geq H(\overline{\mathsf{S}}_{\rightarrow 4}|\mathsf{W}_{[1:3]})+H(\underline{\mathsf{S}}_{\rightarrow 4}).\label{eq:BBQ3}
\end{align}
Adding \eqref{eq:BBQ1}--\eqref{eq:BBQ3} completes the proof of \eqref{eq:BBQ}.

To proceed, we shall consider the cases where $\alpha \geq H(\mathsf{S}_{[2:4]\rightarrow 1})$ and $\alpha \leq H(\mathsf{S}_{[2:4]\rightarrow 1})$, separately.

Case 1: $\alpha \geq H(\mathsf{S}_{[2:4]\rightarrow 1})$. In this case, we have
\begin{align*}
&8\alpha+3H(\mathsf{S}_{\rightarrow 1}) \\
& \geq 7\alpha+3H(\mathsf{S}_{\rightarrow 1})+H(\mathsf{S}_{[2:4]\rightarrow 1})\\
& \stackrel{(a)}{\geq} 7\alpha+\sum_{i=2}^{6}H(\overline{\mathsf{S}}_{\rightarrow i}|\mathsf{W}_{[1:i-1]})+H(\underline{\mathsf{S}}_{\rightarrow 4})+H(\mathsf{S}_{[2:4]\rightarrow 1})\\
& = 7\alpha+\sum_{i=2}^{6}H(\overline{\mathsf{S}}_{\rightarrow i}|\mathsf{W}_{[1:i-1]})+H(\mathsf{S}_{[1:3]\rightarrow 4})+H(\mathsf{S}_{[2:4]\rightarrow 1})\\
& \stackrel{(b)}{=} 7\alpha+\sum_{i=2}^{6}H(\overline{\mathsf{S}}_{\rightarrow i}|\mathsf{W}_{[1:i-1]})+H(\mathsf{S}_{[5:7]\rightarrow 1})+H(\mathsf{S}_{[2:4]\rightarrow 1})\\
& \geq 7\alpha+\sum_{i=2}^{6}H(\overline{\mathsf{S}}_{\rightarrow i}|\mathsf{W}_{[1:i-1]})+H(\mathsf{S}_{\rightarrow 1})\\
& = 7\alpha+\sum_{i=1}^{6}H(\overline{\mathsf{S}}_{\rightarrow i}|\mathsf{W}_{[1:i-1]})\\
& \stackrel{(c)}{\geq} 3H(\mathsf{S}_{\rightarrow 1})+3B
\end{align*}
where $(a)$ follows from \eqref{eq:BBQ}; $(b)$ follows from the fact that $H(\mathsf{S}_{[1:3]\rightarrow 4})=H(\mathsf{S}_{[5:7]\rightarrow 1})$ due to the symmetrical code that we consider; and $(c)$ follows from Lemma~\ref{lemma4} with $n=7$. Canceling $3H(\mathsf{S}_{\rightarrow 1})$ from both sides and normalizing the remaining terms by $B$ complete the proof of \eqref{eq:C4} for the cases where $\alpha \geq H(\mathsf{S}_{[2:4]\rightarrow 1})$.

Case 2: $\alpha \leq H(\mathsf{S}_{[2:4]\rightarrow 1})$. Note that in this case, by node-capacity constraint and the symmetry of the code that we consider, we have
\begin{align}
H(\mathsf{W}_1) \leq \alpha \leq H(\mathsf{S}_{[2:4]\rightarrow 1})= H(\mathsf{S}_{[1:3]\rightarrow 4})=H(\underline{\mathsf{S}}_{\rightarrow 4}).
\label{eq:JH1}
\end{align}
It follows that
\begin{align}
&\sum_{i=2}^{6}H(\overline{\mathsf{S}}_{\rightarrow i}|\mathsf{W}_{[1:i-1]})+H(\underline{\mathsf{S}}_{\rightarrow 4})-H(\mathsf{W}_{[1:6]})\nonumber\\
& =\sum_{i=2}^{6}\left(H(\overline{\mathsf{S}}_{\rightarrow i}|\mathsf{W}_{[1:i-1]})-H(\mathsf{W}_i|\mathsf{W}_{[1:i-1]})\right)\nonumber\\
& \hspace{20pt}+\left(H(\underline{\mathsf{S}}_{\rightarrow 4})-H(\mathsf{W}_1)\right)\nonumber\\
& \stackrel{(a)}{\geq}\sum_{i=2}^{6}\left(H(\overline{\mathsf{S}}_{\rightarrow i}|\mathsf{W}_{[1:i-1]})-H(\mathsf{W}_i|\mathsf{W}_{[1:i-1]})\right)\nonumber\\
& \stackrel{(b)}{=}\sum_{i=2}^{6}\left(H(\overline{\mathsf{S}}_{\rightarrow i},\mathsf{W}_i|\mathsf{W}_{[1:i-1]})-H(\mathsf{W}_i|\mathsf{W}_{[1:i-1]})\right)\nonumber\\
& =\sum_{i=2}^{6}H(\overline{\mathsf{S}}_{\rightarrow i}|\mathsf{W}_{[1:i]})\nonumber\\
& \geq\sum_{i=2}^{5}H(\overline{\mathsf{S}}_{\rightarrow i}|\mathsf{W}_{[1:i]})\nonumber\\
& =\sum_{i=2}^{5}H(\mathsf{S}_{[i+1:7]\rightarrow i}|\mathsf{W}_{[1:i]})\nonumber\\
& \geq\sum_{i=2}^{5}H(\mathsf{S}_{[i+1:6]\rightarrow i}|\mathsf{W}_{[1:i]})\nonumber\\
& = \sum_{i=2}^{5}\sum_{j=i+1}^{6}H(\mathsf{S}_{j\rightarrow i}|\mathsf{W}_{[1:i]},\mathsf{S}_{[i+1:j-1]\rightarrow i})\nonumber\\
& \stackrel{(c)}{\geq} \sum_{i=2}^{5}\sum_{j=i+1}^{6}H(\mathsf{S}_{j\rightarrow i}|\mathsf{W}_{[1:j-1]})\nonumber\\
& = \sum_{j=3}^{6}\sum_{i=2}^{j-1}H(\mathsf{S}_{j\rightarrow i}|\mathsf{W}_{[1:j-1]})\nonumber\\
& \geq \sum_{j=3}^{6}H(\mathsf{S}_{j\rightarrow [2:j-1]}|\mathsf{W}_{[1:j-1]})\nonumber\\
& = \sum_{j=3}^{6}H(\mathsf{S}_{j\rightarrow [2:j-1]})-\sum_{j=3}^{6}I(\mathsf{S}_{j\rightarrow [2:j-1]};\mathsf{W}_{[1:j-1]})\nonumber\\
& \stackrel{(d)}{\geq} \sum_{j=3}^{6}H(\mathsf{S}_{j\rightarrow [2:j-1]})-\sum_{j=3}^{6}I(\mathsf{W}_j;\mathsf{W}_{[1:j-1]})\label{eq:JH2}
\end{align}
where $(a)$ follows from \eqref{eq:JH1}; $(b)$ follows from the fact that $\mathsf{W}_i$ is a function of $(\overline{\mathsf{S}}_{\rightarrow i},\mathsf{W}_{[1:i-1]})=\mathsf{L}_{i-1,i}$ by Lemma~\ref{lemma1}; $(c)$ follows from the fact that $\mathsf{S}_{[i+1:j-1]\rightarrow i}$ is a function of $\mathsf{W}_{[1:j-1]}$; and $(d)$ follows from the fact that  $\mathsf{S}_{j\rightarrow [2:j-1]}$ is a function of $\mathsf{W}_j$. Further note that
\begin{align}
& \sum_{j=3}^{6}H(\mathsf{S}_{j\rightarrow [2:j-1]})+2\alpha \stackrel{(a)}{=} \sum_{j=3}^{6}H(\mathsf{S}_{j+1\rightarrow [3:j]})+2\alpha\nonumber\\
& = \sum_{j=4}^{7}H(\mathsf{S}_{j\rightarrow [3:j-1]})+2\alpha \geq H(\overline{\mathsf{S}}_{\rightarrow [3:6]})+2\alpha\nonumber\\
& \stackrel{(b)}{\geq} H(\overline{\mathsf{S}}_{\rightarrow [3:6]})+H(\mathsf{W}_1)+H(\mathsf{W}_2)\nonumber\\
& \geq H(\overline{\mathsf{S}}_{\rightarrow [3:6]},\mathsf{W}_{[1:2]})=H(\mathsf{L}_{2,6})\label{eq:JH3}
\end{align}
where $(a)$ follows from the fact that $H(\mathsf{S}_{j\rightarrow [2:j-1]})=H(\mathsf{S}_{j+1\rightarrow [3:j]})$ due to the symmetrical code that we consider; and $(b)$ is due to the node-capacity constraint. We thus have
\begin{align*}
&8\alpha+3H(\mathsf{S}_{\rightarrow 1})\\
& \stackrel{(a)}{\geq} 8\alpha+\sum_{i=2}^{6}H(\overline{\mathsf{S}}_{\rightarrow i}|\mathsf{W}_{[1:i-1]})+H(\underline{\mathsf{S}}_{\rightarrow 4})\\
& \stackrel{(b)}{\geq} 8\alpha+\sum_{j=3}^{6}H(\mathsf{S}_{j\rightarrow [2:j-1]})+H(\mathsf{W}_{[1:6]})-\sum_{j=3}^{6}I(\mathsf{W}_j;\mathsf{W}_{[1:j-1]})\\
& \stackrel{(c)}{\geq} 6\alpha+H(\mathsf{L}_{2,6})+H(\mathsf{W}_{[1:6]})-\sum_{j=3}^{6}I(\mathsf{W}_j;\mathsf{W}_{[1:j-1]})\\
& \stackrel{(d)}{\geq} \sum_{j=1}^{6}H(\mathsf{W}_j)+H(\mathsf{L}_{2,6})+H(\mathsf{W}_{[1:6]})-\sum_{j=3}^{6}I(\mathsf{W}_j;\mathsf{W}_{[1:j-1]})\\
& = \sum_{j=2}^{6}I(\mathsf{W}_j;\mathsf{W}_{[1:j-1]})+H(\mathsf{L}_{2,6})+2H(\mathsf{W}_{[1:6]})\nonumber\\
& \hspace{20pt} -\sum_{j=3}^{6}I(\mathsf{W}_j;\mathsf{W}_{[1:j-1]})\\
& = I(\mathsf{W}_1;\mathsf{W}_2)+H(\mathsf{L}_{2,6})+2H(\mathsf{W}_{[1:6]})\nonumber\\
& \geq H(\mathsf{L}_{2,6})+2H(\mathsf{W}_{[1:6]}) \stackrel{(e)}{\geq} 3H(\mathsf{W}_{[1:6]})\stackrel{(f)}{=}3H(\mathsf{W}_{[1:7]},\mathsf{M})\nonumber\\
& \stackrel{(g)}{=} 3H(\mathsf{W}_{[1:7]},\mathsf{M},\mathsf{S}_{\rightarrow 1}) \geq 3H(\mathsf{M},\mathsf{S}_{\rightarrow 1})\\
&= 3H(\mathsf{S}_{\rightarrow 1})+3H(\mathsf{M}|\mathsf{S}_{\rightarrow 1})\stackrel{(h)}{=}3H(\mathsf{S}_{\rightarrow 1})+3H(\mathsf{M})\\
& =3H(\mathsf{S}_{\rightarrow 1})+3B
\end{align*}
where $(a)$ follows from \eqref{eq:BBQ}; $(b)$ follows from \eqref{eq:JH2}; $(c)$ follows from \eqref{eq:JH3}; $(d)$ follows from the node-capacity constraint; $(e)$ follows from the fact that $H(\mathsf{L}_{2,6})\geq H(\mathsf{W}_{[1:6]})$ due to Lemma~\ref{lemma1}; $(f)$ follows from the facts that $\mathsf{M}$ is a function of $\mathsf{W}_{[1:6]}$ due to the message-recovery requirement \eqref{eq:Cons2} and that $\mathsf{W}_7$ is a function of $\mathsf{S}_{\rightarrow 7}$, which is in turn a function of $\mathsf{W}_{[1:6]}$; $(g)$ follows from the fact that $\mathsf{S}_{\rightarrow 1}$ is a function of $\mathsf{W}_{[2:7]}$; and $(h)$ follows from the repair-secrecy constraint \eqref{eq:Cons4} with $\ell=1$. Canceling $3H(\mathsf{S}_{\rightarrow 1})$ from both sides and normalizing the remaining terms by $B$ complete the proof of \eqref{eq:C4} for the cases where $\alpha \geq H(\mathsf{S}_{[2:4]\rightarrow 1})$.

\section{Concluding Remarks}
In this paper, we considered the $(n,k,d,\ell)$ secure exact-repair regenerating code problem, which has been previously studied in \cite{PRR-ISIT10,SRK-Globecom11,TACB-IT16,YSY-ISIT16}. We proved that when the secrecy parameter $\ell$ is sufficiently large, the SRK point \cite{SRK-Globecom11} is the {\em only} corner point of the achievable normalized storage-capacity repair-bandwidth tradeoff region. This includes all previous results from \cite{TACB-IT16} and \cite{YSY-ISIT16} as special cases. On the other hand, when $\ell$ is small, we showed that it is entirely possible that the achievable normalized storage-capacity repair-bandwidth tradeoff region features {\em multiple} corner points. In particular, we showed that the achievable normalized storage-capacity repair-bandwidth tradeoff region for the $(7,6,6,1)$ problem has exactly {\em two} corner points. This suggests a much ``smoother" transition, in terms of the rate region, from the original exact-repair regenerating code problem to the secrecy extension than that suggested by the previous results from \cite{TACB-IT16} and \cite{YSY-ISIT16}.

The question whether \eqref{eq:Cond} is also {\em necessary} for the SRK point \cite{SRK-Globecom11} to be the {\em only} corner point of the achievable normalized storage-capacity repair-bandwidth tradeoff region remains open. Significant research is also needed to further understand how the tradeoff region $\mathcal{R}_{n,k,d,\ell}$ may look like when $\ell$ is small (the non-secrecy case with $\ell=0$ remains open and appears to be very challenging).

\section*{Acknowledgment}
The authors would like to thank Dr. Kenneth Shum for sharing with us an early draft of his paper \cite{YSY-ISIT16}, which inspired our interest in the secure exact-repair regenerating code problem. 

\appendix[Proof of the Technical Lemmas]

{\em Proof of Lemma~\ref{lemma1}.} Fix $s \in [1:n]$ and $t\in[0:s-1]$. Let us first note that $\underline{\mathsf{S}}_{\rightarrow t+1}$ is a function of $\mathsf{W}_{[1:t]}$. As a result, $\mathsf{S}_{\rightarrow t+1}=(\underline{\mathsf{S}}_{\rightarrow t+1},\overline{\mathsf{S}}_{\rightarrow t+1})$ is a function of $\mathsf{L}_{t,s}$. It thus follows immediately from the node-regeneration requirement \eqref{eq:Cons3} that $\mathsf{W}_{t+1}$ is a function of $\mathsf{L}_{t,s}$. Similarly and inductively, it can be shown that $(\underline{\mathsf{S}}_{\rightarrow j},\mathsf{W}_j)$ is a function of $\mathsf{L}_{t,s}$ for all $j\in [t+2:s]$ as well. This completes the proof of Lemma~\ref{lemma1}.

{\em Proof of Lemma~\ref{lemma2}.} To prove \eqref{eq:LLL}, let us fix $t\in[1:2]$, $r\in[2:k-1]$, $p\in[1:r-t+1]$, and $q\in[0:d-r-1]$. We have
\begin{align*}
&H(\mathsf{S}_{1\rightarrow [2:p+1]})+H(\mathsf{L}_{t,r},\mathsf{S}_{[r+2:r+q+1]\rightarrow r+1})\\
&\stackrel{(a)}{=} H(\mathsf{S}_{r+q+2\rightarrow [r-p+2:r+1]})+H(\mathsf{L}_{t,r},\mathsf{S}_{[r+2:r+q+1]\rightarrow r+1})\\
&\stackrel{(b)}{\geq} H(\mathsf{S}_{r+q+2\rightarrow [r-p+2:r]})+H(\mathsf{L}_{t,r},\mathsf{S}_{[r+2:r+q+2]\rightarrow r+1})\\
&\stackrel{(c)}{=}H(\mathsf{S}_{1\rightarrow [2:p]})+H(\mathsf{L}_{t,r},\mathsf{S}_{[r+2:r+q+2]\rightarrow r+1})
\end{align*}
where $(a)$ follows from the fact that $H(\mathsf{S}_{1\rightarrow [2:p+1]})=H(\mathsf{S}_{r+q+2\rightarrow [r-p+2:r+1]})$ due to the symmetrical codes that we consider; $(b)$ follows from the submodularity of the entropy function; and $(c)$ follows from the fact that $H(\mathsf{S}_{r+q+2\rightarrow [r-p+2:r]})=H(\mathsf{S}_{1\rightarrow [2:p]})$ again due to the symmetrical codes that we consider. This completes the proof of \eqref{eq:LLL} for any $r\in[2:k-1]$, $p\in[1:r-t+1]$, and $q\in[0:d-r-1]$.

To prove \eqref{eq:P3}, let us fix $t\in[1:2]$, $j\in[2:k]$, and $m\in[1:j-t+1]$. Notice that \eqref{eq:P3} holds trivially with equality when $j=k$, so we only need to consider the cases where $j\in[2:k-1]$ for $k \geq 3$. (When $k=2$, $[2:k-1]$ is empty and there is nothing to prove.) Now adding \eqref{eq:LLL} for $q\in[0:d-r-1]$ gives:
\begin{align*}
&(d-r)H(\mathsf{S}_{1\rightarrow [2:p+1]})+\sum_{q=0}^{d-r-1}H(\mathsf{L}_{t,r},\mathsf{S}_{[r+2:r+q+1]\rightarrow r+1})\nonumber\\
& \geq (d-r)H(\mathsf{S}_{1\rightarrow [2:p]})+\sum_{q=0}^{d-r-1}H(\mathsf{L}_{t,r},\mathsf{S}_{[r+2:r+q+2]\rightarrow r+1}).
\end{align*}
Canceling $\sum_{q=1}^{d-r-1}H(\mathsf{L}_{t,r},\mathsf{S}_{[r+2:r+q+1]\rightarrow r+1})$ from both sides of the above inequality gives:
\begin{align}
&(d-r)H(\mathsf{S}_{1\rightarrow [2:p+1]})+H(\mathsf{L}_{t,r})\nonumber\\
& \hspace{10pt} \geq (d-r)H(\mathsf{S}_{1\rightarrow [2:p]})+H(\mathsf{L}_{t,r},\mathsf{S}_{[r+2:n]\rightarrow r+1})\nonumber\\
& \hspace{10pt} = (d-r)H(\mathsf{S}_{1\rightarrow [2:p]})+H(\mathsf{L}_{t,r+1})\label{eq:LLL1}
\end{align}
for any $r\in[2:k-1]$ and $p\in[1:r-t+1]$. Adding \eqref{eq:LLL1} for $r\in[j:k-1]$ and $p\in[1:m]$ gives:
\begin{align*}
&T_{k,d,j}\sum_{p=1}^{m}H(\mathsf{S}_{1\rightarrow [2:p+1]})+m\sum_{r=j}^{k-1}H(\mathsf{L}_{t,r})\nonumber\\
& \hspace{10pt} \geq T_{k,d,j}\sum_{p=1}^{m}H(\mathsf{S}_{1\rightarrow [2:p]})+m\sum_{r=j}^{k-1}H(\mathsf{L}_{t,r+1}).
\end{align*}
Canceling $T_{k,d,j}\sum_{p=2}^{m}H(\mathsf{S}_{1\rightarrow [2:p]})+m\sum_{r=j+1}^{k-1}H(\mathsf{L}_{t,r})$ from both sides of the above inequality gives:
\begin{align*}
T_{k,d,j}H(\mathsf{S}_{1\rightarrow [2:m+1]})+mH(\mathsf{L}_{t,j}) \geq mH(\mathsf{L}_{t,k}).
\end{align*}
Dividing both sides by $m$ completes the proof of \eqref{eq:P3} for any $t\in[1:2]$, $j\in[2:k]$, and $m\in[1:j-t+1]$. This completes the proof of Lemma~\ref{lemma2}.

{\em Proof of Lemma~\ref{lemma3}.} To prove \eqref{eq:exchange3}, let us fix $j\in[2:k-1]$ and $t\in [j:k-1]$. Let $n-j=u(d-t)+p$ for some positive integers $u$ and $p \in [1:d-t]$. Let
\begin{align}
\tau_0:=\{1\}\cup[j+1:j+p-1]
\end{align}
and
\begin{align}
\tau_q:=[j+p+(q-1)(d-t):j+p+q(d-t)-1]
\end{align}
for $q\in[1:u-1]$. Notice that we have
\begin{align}
\cup_{q=0}^{u-1}\tau_q = \{1\}\cup[j+1:t].\label{eq:Tau}
\end{align}

By the symmetry of the codes that we consider, we have
\begin{align}
H(\mathsf{S}_{\tau_0\rightarrow t+1}&|\mathsf{W}_2,\mathsf{S}_{\rightarrow [3:j]},\mathsf{S}_{t+1,2})\nonumber\\
&=H(\mathsf{S}_{B\rightarrow t+1}|\mathsf{W}_2,\mathsf{S}_{\rightarrow [3:j]},\mathsf{S}_{t+1,2})
\end{align}
for any $B\subseteq [t+2:n]$ such that $|B|=|\tau_0|=p$. It follows that 
\begin{align}
H&(\mathsf{S}_{\tau_0\rightarrow t+1}|\mathsf{W}_2,\mathsf{S}_{\rightarrow [3:j]},\mathsf{S}_{t+1\rightarrow 2})\nonumber\\ 
& \stackrel{(a)}{\geq} \frac{p}{d-t}H(\mathsf{S}_{[t+2:n]\rightarrow t+1}|\mathsf{W}_2,\mathsf{S}_{\rightarrow [3:j]},\mathsf{S}_{t+1\rightarrow 2})\nonumber\\
& = \frac{p}{d-t}H(\mathsf{S}_{[t+2:n]\rightarrow t+1}|\mathsf{W}_2,\underline{\mathsf{S}}_{\rightarrow [3:j]},\overline{\mathsf{S}}_{\rightarrow [3:j]},\mathsf{S}_{t+1\rightarrow 2})\nonumber\\
& \stackrel{(b)}{\geq} \frac{p}{d-t}H(\mathsf{S}_{[t+2:n]\rightarrow t+1}|\mathsf{W}_2,\underline{\mathsf{S}}_{\rightarrow [3:j]},\mathsf{L}_{1,t})\nonumber\\
& \stackrel{(c)}{=} \frac{p}{d-t}H(\mathsf{S}_{[t+2:n]\rightarrow t+1}|\mathsf{L}_{1,t})\nonumber\\
& = \frac{p}{d-t}\left(H(\mathsf{S}_{[t+2:n]\rightarrow t+1},\mathsf{L}_{1,t})-H(\mathsf{L}_{1,t})\right)\nonumber\\
& = \frac{p}{d-t}\left(H(\mathsf{L}_{1,t+1})-H(\mathsf{L}_{1,t})\right)\label{eq:Tau0}
\end{align}
where $(a)$ follows from the well-known Han's inequality \cite{Han-IC78}; $(b)$ follows from the facts that $(\overline{\mathsf{S}}_{\rightarrow [3:j]},\mathsf{S}_{t+1\rightarrow 2})$ is a sub-collection of random variables from $\mathsf{L}_{1,t}$ and that conditioning reduces entropy; and $(c)$ follows from the fact that $(\mathsf{W}_2,\underline{\mathsf{S}}_{\rightarrow [3:j]})$ is a function of $\mathsf{L}_{1,t}$ by Lemma~\ref{lemma1}.

Next, let us show, by induction, that 
\begin{align}
&rH(\mathsf{L}_{1,t})+H(\mathsf{L}_{1,j},\mathsf{S}_{j\rightarrow 1})\nonumber\\
& \geq rH(\mathsf{L}_{1,t+1})+H(\mathsf{W}_2,\mathsf{S}_{\rightarrow [3:j]},\mathsf{S}_{\cup_{q=0}^{u-r}\tau_q\rightarrow t+1},\mathsf{S}_{t+1\rightarrow 2})\label{eq:IND}
\end{align}
for any $r\in[1:u]$. 

For the base case where $r=1$, we have
\begin{align*}
&H(\mathsf{L}_{1,t})+H(\mathsf{L}_{1,j},\mathsf{S}_{j\rightarrow 1})\\
&\stackrel{(a)}{=}H(\mathsf{L}_{1,t},\mathsf{W}_2,\underline{\mathsf{S}}_{\rightarrow[2:j]},\mathsf{S}_{\cup_{q=0}^{u-1}\tau_q\rightarrow t+1})+H(\mathsf{L}_{1,j},\mathsf{S}_{j\rightarrow 1})\\
&\stackrel{(b)}{=}H(\mathsf{L}_{1,t},\mathsf{W}_2,\underline{\mathsf{S}}_{\rightarrow[2:j]},\mathsf{S}_{\cup_{q=0}^{u-1}\tau_q\rightarrow t+1})\\
&\hspace{10pt}+H(\mathsf{L}_{1,j},\underline{\mathsf{S}}_{\rightarrow[2:j-1]},\mathsf{S}_{j\rightarrow 1})\\
&=H(\mathsf{W}_{[1:2]},\mathsf{S}_{\rightarrow[2:j]},\overline{\mathsf{S}}_{\rightarrow [j+1:t]},\mathsf{S}_{\cup_{q=0}^{u-1}\tau_q\rightarrow t+1})\\
&\hspace{10pt}+H(\mathsf{W}_1,\mathsf{S}_{\rightarrow[2:j-1]},\overline{\mathsf{S}}_{\rightarrow j},\mathsf{S}_{j\rightarrow 1})\\
&\stackrel{(c)}{=} H(\mathsf{W}_{[1:2]},\mathsf{S}_{\rightarrow[2:j]},\overline{\mathsf{S}}_{\rightarrow [j+1:t]},\mathsf{S}_{\cup_{q=0}^{u-1}\tau_q\rightarrow t+1})\\
& \hspace{10pt}+H(\mathsf{W}_2,\mathsf{S}_{\rightarrow [3:j]},\mathsf{S}_{\cup_{q=0}^{u-1}\tau_q\rightarrow t+1},\overline{\mathsf{S}}_{\rightarrow t+1},\mathsf{S}_{t+1\rightarrow 2})\\
&\stackrel{(d)}{\geq} H(\mathsf{W}_{[1:2]},\mathsf{S}_{\rightarrow [2:j]}, \overline{\mathsf{S}}_{\rightarrow [j+1:t+1]},\mathsf{S}_{\cup_{q=0}^{u-1}\tau_q\rightarrow t+1})\\
& \hspace{10pt}+H(\mathsf{W}_2,\mathsf{S}_{\rightarrow [3:j]},\mathsf{S}_{\cup_{q=0}^{u-1}\tau_q\rightarrow t+1},\mathsf{S}_{t+1\rightarrow 2})\\
& =H(\mathsf{L}_{1,t+1},\mathsf{W}_2,\underline{\mathsf{S}}_{\rightarrow [2:j]},\mathsf{S}_{\cup_{q=0}^{u-1}\tau_q\rightarrow t+1})\\
& \hspace{10pt}+H(\mathsf{W}_2,\mathsf{S}_{\rightarrow [3:j]},\mathsf{S}_{\cup_{q=0}^{u-1}\tau_q\rightarrow t+1},\mathsf{S}_{t+1\rightarrow 2})\\
&\stackrel{(e)}{=} H(\mathsf{L}_{1,t+1})+H(\mathsf{W}_2,\mathsf{S}_{\rightarrow [3:j]},\mathsf{S}_{\cup_{q=0}^{u-1}\tau_q\rightarrow t+1},\mathsf{S}_{t+1\rightarrow 2})
\end{align*}
where $(a)$ and $(e)$ follow from the fact that $(\mathsf{W}_2,\underline{\mathsf{S}}_{\rightarrow[2:j]},\mathsf{S}_{\cup_{q=0}^{u-r}\tau_q\rightarrow t+1})$ is a function of $\mathsf{L}_{1,t}$ by Lemma~\ref{lemma1}; $(b)$ follows from the fact that $\underline{\mathsf{S}}_{\rightarrow[2:j]}$ is a function of $\mathsf{L}_{1,j}$ by Lemma~\ref{lemma1}; and $(c)$ follows from the fact that 
\begin{align}
&H(\mathsf{W}_1,\mathsf{S}_{\rightarrow[2:j-1]},\overline{\mathsf{S}}_{\rightarrow j},\mathsf{S}_{j\rightarrow 1})
\nonumber\\
&=H(\mathsf{W}_2,\mathsf{S}_{\rightarrow [3:j]},\mathsf{S}_{\cup_{q=0}^{u-1}\tau_q\rightarrow t+1},\overline{\mathsf{S}}_{\rightarrow t+1},\mathsf{S}_{t+1\rightarrow 2})
\end{align}
due to the symmetrical codes that we consider; and $(d)$ follows from the submodularity of the entropy function. This completes the proof of the base case.

Now assume that \eqref{eq:IND} holds for some $r\in [1:u-1]$. Similar to the base case, we have
\begin{align}
&H(\mathsf{L}_{1,t})+H(\mathsf{W}_2,\mathsf{S}_{\rightarrow [3:j]},\mathsf{S}_{\cup_{q=0}^{u-r}\tau_q\rightarrow t+1},\mathsf{S}_{t+1\rightarrow 2})\nonumber\\
& \stackrel{(a)}{=} H(\mathsf{L}_{1,t},\mathsf{W}_2,\underline{\mathsf{S}}_{\rightarrow[2:j]},\mathsf{S}_{\cup_{q=0}^{u-(r+1)}\tau_q\rightarrow t+1})\nonumber\\
& \hspace{10pt} +H(\mathsf{W}_2,\mathsf{S}_{\rightarrow [3:j]},\mathsf{S}_{\cup_{q=0}^{u-r}\tau_q\rightarrow t+1},\mathsf{S}_{t+1\rightarrow 2})\nonumber\\
& = H(\mathsf{W}_{[1:2]},\mathsf{S}_{\rightarrow [2:j]},\overline{\mathsf{S}}_{\rightarrow[j+1:t]},\mathsf{S}_{\cup_{q=0}^{u-(r+1)}\tau_q\rightarrow t+1})\nonumber\\
& \hspace{10pt} +H(\mathsf{W}_2,\mathsf{S}_{\rightarrow [3:j]},\mathsf{S}_{\cup_{q=0}^{u-r}\tau_q\rightarrow t+1},\mathsf{S}_{t+1\rightarrow 2})\nonumber\\
& \stackrel{(b)}{=} H(\mathsf{W}_{[1:2]},\mathsf{S}_{\rightarrow [2:j]},\overline{\mathsf{S}}_{\rightarrow[j+1:t]},\mathsf{S}_{\cup_{q=0}^{u-(r+1)}\tau_q\rightarrow t+1})\nonumber\\
& \hspace{10pt} +H(\mathsf{W}_2,\mathsf{S}_{\rightarrow [3:j]},\mathsf{S}_{\cup_{q=0}^{u-(r+1)}\tau_q\rightarrow t+1},\overline{\mathsf{S}}_{\rightarrow t+1},\mathsf{S}_{t+1\rightarrow 2})\nonumber\\
& \stackrel{(c)}{\geq} H(\mathsf{W}_{[1:2]},\mathsf{S}_{\rightarrow [2:j]},\overline{\mathsf{S}}_{\rightarrow [j+1:t+1]},\mathsf{S}_{\cup_{q=0}^{u-(r+1)}\tau_q\rightarrow t+1})\nonumber\\
& \hspace{10pt} +H(\mathsf{W}_2,\mathsf{S}_{\rightarrow [3:j]},\mathsf{S}_{\cup_{q=0}^{u-(r+1)}\tau_q\rightarrow t+1},\mathsf{S}_{t+1\rightarrow 2})\nonumber\\
& = H(\mathsf{L}_{1,t+1},\mathsf{W}_2,\underline{\mathsf{S}}_{\rightarrow [2:j]},\mathsf{S}_{\cup_{q=0}^{u-(r+1)}\tau_q\rightarrow t+1})\nonumber\\
& \hspace{10pt} +H(\mathsf{W}_2,\mathsf{S}_{\rightarrow [3:j]},\mathsf{S}_{\cup_{q=0}^{u-(r+1)}\tau_q\rightarrow t+1},\mathsf{S}_{t+1\rightarrow 2})\nonumber\\
& \stackrel{(d)}{=} H(\mathsf{L}_{1,t+1})+H(\mathsf{W}_2,\mathsf{S}_{\rightarrow [3:j]},\mathsf{S}_{\cup_{q=0}^{u-(r+1)}\tau_q\rightarrow t+1},\mathsf{S}_{t+1\rightarrow 2})\label{eq:IND2}
\end{align}
where $(a)$ and $(d)$ follow from the fact that $(\mathsf{W}_2,\underline{\mathsf{S}}_{\rightarrow[2:j]},\mathsf{S}_{\cup_{q=0}^{u-(r+1)}\tau_q\rightarrow t+1})$ is a function of $\mathsf{L}_{1,t+1}$ by Lemma~\ref{lemma1}; $(b)$ follows from the fact that
\begin{align}
&H(\mathsf{W}_2,\mathsf{S}_{\rightarrow [3:j]},\mathsf{S}_{\cup_{q=0}^{u-r}\tau_q\rightarrow t+1},\mathsf{S}_{t+1\rightarrow 2})\nonumber\\
&=H(\mathsf{W}_2,\mathsf{S}_{\rightarrow [3:j]},\mathsf{S}_{\cup_{q=0}^{u-(r+1)}\tau_q\rightarrow t+1},\overline{\mathsf{S}}_{\rightarrow t+1},\mathsf{S}_{t+1\rightarrow 2})
\end{align}
due to the symmetrical codes that we consider; and $(c)$ follows from the submodularity of the entropy function. Adding \eqref{eq:IND} and \eqref{eq:IND2} gives
\begin{align*}
(r+&1)H(\mathsf{L}_{1,t})+H(\mathsf{L}_{1,j},\mathsf{S}_{j\rightarrow 1})\\
& \geq (r+1)H(\mathsf{L}_{1,t+1})\\
& \hspace{10pt}+H(\mathsf{W}_2,\mathsf{S}_{\rightarrow [3:j]},\mathsf{S}_{\cup_{q=0}^{u-(r+1)}\tau_q\rightarrow t+1},\mathsf{S}_{t+1\rightarrow 2}).
\end{align*}
This completes the proof of the induction step.

Finally, setting $r=u$ in \eqref{eq:IND} gives:
\begin{align}
&uH(\mathsf{L}_{1,t})+H(\mathsf{L}_{1,j},\mathsf{S}_{j\rightarrow 1})\nonumber\\
& \geq uH(\mathsf{L}_{1,t+1})+H(\mathsf{W}_2,\mathsf{S}_{\rightarrow [3:j]},\mathsf{S}_{\tau_0\rightarrow t+1},\mathsf{S}_{t+1\rightarrow 2})\nonumber\\
& = uH(\mathsf{L}_{1,t+1})+H(\mathsf{W}_2,\mathsf{S}_{\rightarrow [3:j]},\mathsf{S}_{t+1\rightarrow 2})\nonumber\\
&\hspace{10pt}+H(\mathsf{S}_{\tau_0\rightarrow t+1}|\mathsf{W}_2,\mathsf{S}_{\rightarrow [3:j]},\mathsf{S}_{t+1\rightarrow 2}).\label{eq:IND3}
\end{align}
Substituting \eqref{eq:Tau0} into \eqref{eq:IND3} and using the fact that
\begin{align}
u+\frac{p}{d-t}=\frac{u(d-t)+p}{d-t}=\frac{n-j}{d-t}
\end{align}
we have
\begin{align}
&\frac{n-j}{d-t}H(\mathsf{L}_{1,t})+H(\mathsf{L}_{1,j},\mathsf{S}_{j\rightarrow 1})\nonumber\\
& \geq \frac{n-j}{d-t}H(\mathsf{L}_{1,t+1})+H(\mathsf{W}_2,\mathsf{S}_{\rightarrow [3:j]},\mathsf{S}_{t+1\rightarrow 2}).
\label{eq:IND4}
\end{align}
Finally, due to the symmetrical codes that we consider, we have
\begin{align}
H&(\mathsf{W}_2,\mathsf{S}_{\rightarrow [3:j]},\mathsf{S}_{t+1\rightarrow 2})\nonumber\\
&=H(\mathsf{W}_1,\mathsf{S}_{\rightarrow [2:j-1]},\mathsf{S}_{j\rightarrow 1})\nonumber\\
&=H(\mathsf{W}_1,\overline{\mathsf{S}}_{\rightarrow [2:j-1]},\underline{\mathsf{S}}_{\rightarrow [2:j-1]},\mathsf{S}_{j\rightarrow 1})\nonumber\\
&=H(\mathsf{L}_{1,j-1},\underline{\mathsf{S}}_{\rightarrow [2:j-1]},\mathsf{S}_{j\rightarrow 1})\nonumber\\
&=H(\mathsf{L}_{1,j-1},\mathsf{S}_{j\rightarrow 1})\label{eq:IND5}
\end{align}
where the last equality follows from the fact that $\underline{\mathsf{S}}_{\rightarrow [2:j-1]}$ is a function of $\mathsf{L}_{1,j-1}$ by Lemma~\ref{lemma1}. Substituting \eqref{eq:IND5} into \eqref{eq:IND4} completes the proof of \eqref{eq:exchange3} for any $j\in[2:k-1]$ and $t\in [j:k-1]$.

{\em Proof of Lemma~\ref{lemma4}.} First note that
\begin{align}
& \sum_{i=1}^{n-1}H(\overline{\mathsf{S}}_{\rightarrow i}|\mathsf{W}_{[1:i-1]})\nonumber\\
& \stackrel{(a)}{=} \sum_{i=1}^{n-1}H(\overline{\mathsf{S}}_{\rightarrow i},\mathsf{W}_i|\mathsf{W}_{[1:i-1]})\nonumber\\
& = \sum_{i=1}^{n-1}H(\overline{\mathsf{S}}_{\rightarrow i}|\mathsf{W}_{[1:i]})+\sum_{i=1}^{n-1}H(\mathsf{W}_i|\mathsf{W}_{[1:i-1]})\nonumber\\
& = \sum_{i=1}^{n-1}H(\overline{\mathsf{S}}_{\rightarrow i}|\mathsf{W}_{[1:i]})+H(\mathsf{W}_{[1:n-1]})\nonumber\\
& = \sum_{i=1}^{n-1}H(\mathsf{S}_{[i+1:n]\rightarrow i}|\mathsf{W}_{[1:i]})+H(\mathsf{W}_{[1:n-1]})\nonumber\\
& = \sum_{i=1}^{n-1}\sum_{j=i+1}^{n}H(\mathsf{S}_{j\rightarrow i}|\mathsf{W}_{[1:i]},\mathsf{S}_{[i+1;j-1]\rightarrow i})+H(\mathsf{W}_{[1:n-1]})\nonumber\\
& \stackrel{(b)}{\geq} \sum_{i=1}^{n-1}\sum_{j=i+1}^{n}H(\mathsf{S}_{j\rightarrow i}|\mathsf{W}_{[1:j-1]})+H(\mathsf{W}_{[1:n-1]})\nonumber\\
& = \sum_{j=2}^{n}\sum_{i=1}^{j-1}H(\mathsf{S}_{j\rightarrow i}|\mathsf{W}_{[1:j-1]})+H(\mathsf{W}_{[1:n-1]})\nonumber\\
& \geq \sum_{j=2}^{n-1}\sum_{i=1}^{j-1}H(\mathsf{S}_{j\rightarrow i}|\mathsf{W}_{[1:j-1]})+H(\mathsf{W}_{[1:n-1]})\nonumber\\
& \geq \sum_{j=2}^{n-1}H(\mathsf{S}_{j\rightarrow [1:j-1]}|\mathsf{W}_{[1:j-1]})+H(\mathsf{W}_{[1:n-1]})\nonumber\\
& = \sum_{j=2}^{n-1}H(\mathsf{S}_{j\rightarrow [1:j-1]})-\sum_{j=1}^{n-1}I(\mathsf{S}_{j\rightarrow [1:j-1]};\mathsf{W}_{[1:j-1]})\nonumber\\
&\hspace{20pt}+H(\mathsf{W}_{[1:n-1]})\nonumber\\
& \stackrel{(c)}{=} \sum_{j=2}^{n-1}H(\mathsf{S}_{j+1\rightarrow [2:j]})-\sum_{j=1}^{n-1}I(\mathsf{S}_{j\rightarrow [1:j-1]};\mathsf{W}_{[1:j-1]})\nonumber\\
&\hspace{20pt}+H(\mathsf{W}_{[1:n-1]})\nonumber\\
& \geq H(\overline{\mathsf{S}}_{\rightarrow [2:n-1]})-\sum_{j=1}^{n-1}I(\mathsf{S}_{j\rightarrow [1:j-1]};\mathsf{W}_{[1:j-1]})\nonumber\\
&\hspace{20pt}+H(\mathsf{W}_{[1:n-1]})\label{eq:Ph7}
\end{align}
where $(a)$ follows from the fact that $\mathsf{W}_i$ is a function of $(\mathsf{W}_{[1:i-1]},\overline{\mathsf{S}}_{\rightarrow i})=\mathsf{L}_{i-1,i}$ by Lemma~\ref{lemma1}; $(b)$ follows from the fact that $\mathsf{S}_{[i+1;j-1]\rightarrow i}$ is a function of $\mathsf{W}_{[1:j-1]}$; and $(c)$ follows from the fact that $H(\mathsf{S}_{j\rightarrow [1:j-1]})=H(\mathsf{S}_{j+1\rightarrow [2:j]})$ due to the symmetrical code that we consider. 

Further note that
\begin{align}
H(\overline{\mathsf{S}}_{\rightarrow [2:n-1]})&+H(\mathsf{W}_1)\nonumber\\
&\geq H(\mathsf{W}_1,\overline{\mathsf{S}}_{\rightarrow [2:n-1]})=H(\mathsf{L}_{1,n-1}).\label{eq:Ph8}
\end{align}
Adding \eqref{eq:Ph7}--\eqref{eq:Ph8} gives:
\begin{align}
&\sum_{i=1}^{n-1}H(\overline{\mathsf{S}}_{\rightarrow i}|\mathsf{W}_{[1:i-1]})+H(\mathsf{W}_1)\nonumber\\
&\geq H(\mathsf{L}_{1,n-1})+H(\mathsf{W}_{[1:n-1]})-\sum_{j=1}^{n-1}I(\mathsf{S}_{j\rightarrow [1:j-1]};\mathsf{W}_{[1:j-1]})\nonumber\\
& \stackrel{(a)}{\geq} H(\mathsf{L}_{1,n-1})+H(\mathsf{W}_{[1:n-1]})-\sum_{j=1}^{n-1}I(\mathsf{W}_j;\mathsf{W}_{[1:j-1]})\nonumber\\
& = H(\mathsf{L}_{1,n-1})+H(\mathsf{W}_{[1:n-1]})\nonumber\\
& \hspace{20pt} -\sum_{j=1}^{n-1}\left(H(\mathsf{W}_j)+H(\mathsf{W}_{[1:j-1]})-H(\mathsf{W}_{[1:j]})\right)\nonumber\\
& = H(\mathsf{L}_{1,n-1})+2H(\mathsf{W}_{[1:n-1]})-\sum_{j=1}^{n-1}H(\mathsf{W}_j)\nonumber\\
& \stackrel{(b)}{\geq} 3H(\mathsf{W}_{[1:n-1]})-\sum_{j=1}^{n-1}H(\mathsf{W}_j)
\label{eq:Ph9}
\end{align}
where $(a)$ follows from the fact that $\mathsf{S}_{j\rightarrow [1:j-1]}$ is a function of $\mathsf{W}_j$; and $(b)$ follows from the fact that $\mathsf{W}_{[2:n-1]}$ is a function of $\mathsf{L}_{1,n-1}$ by Lemma~\ref{lemma1} so we have $H(\mathsf{L}_{1,n-1}) \geq H(\mathsf{W}_{[1:n-1]})$. 

Finally, by the node-capacity constraint, we have
\begin{align}
n\alpha \geq H(\mathsf{W}_1)+\sum_{j=1}^{n-1}H(\mathsf{W}_j).
\label{eq:Ph10}
\end{align}
Adding \eqref{eq:Ph9} and \eqref{eq:Ph10} gives:
\begin{align*}
&\sum_{i=1}^{n-1}H(\overline{\mathsf{S}}_{\rightarrow i}|\mathsf{W}_{[1:i-1]})+n\alpha\\
& \geq 3H(\mathsf{W}_{[1:n-1]}) \stackrel{(a)}{=} 3H(\mathsf{W}_{[1:n]},\mathsf{M}) \stackrel{(b)}{=} 3H(\mathsf{W}_{[1:n]},\mathsf{M},\mathsf{S}_{\rightarrow 1})\\
& \geq 3H(\mathsf{M},\mathsf{S}_{\rightarrow 1}) = 3H(\mathsf{S}_{\rightarrow 1})+3H(\mathsf{M}|\mathsf{S}_{\rightarrow 1})\\
& \stackrel{(c)}{=} 3H(\mathsf{S}_{\rightarrow 1})+3H(\mathsf{M}) = 3H(\mathsf{S}_{\rightarrow 1})+3B
\end{align*}
where $(a)$ follows from the facts that $\mathsf{M}$ is a function of $\mathsf{W}_{[1:n-1]}$ by the message-recovery requirement \eqref{eq:Cons2} and that $\mathsf{W}_n$ is a function of $\mathsf{S}_{\rightarrow n}$, which is in turn a function of $\mathsf{W}_{[1:n-1]}$; $(b)$ follows from the fact that $\mathsf{S}_{\rightarrow 1}$ is a function of $\mathsf{W}_{[2:n]}$; and $(c)$ follows from the repair-secrecy requirement \eqref{eq:Cons4} with $\ell=1$. This completes the proof of Lemma~\ref{lemma4}.

\bibliographystyle{ieeetr}

\end{document}